\documentclass{aa}
\usepackage[varg]{txfonts}
\usepackage[T1]{fontenc}
\usepackage{graphicx}	
\usepackage{xcolor}	
\usepackage{amsmath}	
\usepackage{amssymb}	
\usepackage{caption}    
\usepackage{subcaption} 
\usepackage{footmisc}
\usepackage{upgreek}
\usepackage{orcidlink}
\usepackage{placeins}
\bibpunct{(}{)}{;}{a}{}{,} 
\usepackage{soul} 
\usepackage{ulem} 
\usepackage{dblfloatfix}
\usepackage{float}
\usepackage{wrapfig}

\usepackage{booktabs}  
\usepackage{longtable}
\usepackage{multicol}  
\usepackage[flushleft]{threeparttable}

\newcommand{\Msun}{{\rm M_{\odot}}}

\newcommand{\kpc}{{\rm kpc}}

\newcommand{\stream}{\rm s}

\begin{document}

\title{The hydrodynamical response of cold circumgalactic clouds to quasar radiation}

\author{Nicolas Ledos\orcid{0000-0001-9699-8941}\inst{1}
    \and Sebastiano Cantalupo\orcid{0000-0001-5804-1428}\inst{1}
    \and Titouan Lazeyras\orcid{0000-0002-7080-9839}\inst{1}
    \and Gabriele Pezzulli\orcid{0000-0003-0736-7879}\inst{2}
    \and Kentaro Nagamine\orcid{0000-0001-7457-8487}\inst{3,4,5,6}
    \and Shinsuke Takasao\orcid{0000-0003-3882-3945}\inst{7}
    \and Marta Galbiati\orcid{0000-0002-1843-1699}\inst{1}
    \and Andrea Travascio\orcid{0000-0002-8863-888X}\inst{8}
    \and Giada Quadri\orcid{0009-0005-9819-5620}\inst{1} 
    \and Weichen Wang\orcid{0000-0002-9593-8274}\inst{1} 
    \and Antonio Pensabene\orcid{0000-0001-9815-4953}\inst{9,10} }   

\offprints{N. Ledos, \email{nicolas.ledos@unimib.it}}

\institute{Dipartimento di Fisica "G. Occhialini”, Universit\`{a} degli Studi di Milano-Bicocca, Piazza della Scienza 3, 20126 Milano, Italy
\and
Kapteyn Astronomical Institute, University of Groningen, NL-9747 AD Groningen, the Netherlands
\and
Theoretical Astrophysics, Department of Earth and Space Science, Graduate School of Science, Osaka University, 1-1 Machikaneyama, Toyonaka, Osaka 560-0043, Japan
\and
Theoretical Joint Research, Forefront Research Center, Graduate School of Science, Osaka University, 1-1 Machikaneyama, Toyonaka, Osaka 560-0043, Japan
\and
Kavli IPMU (WPI), UTIAS, The University of Tokyo, 5-1-5 Kashiwanoha, Kashiwa, Chiba 277-8583, Japan
\and
Department of Physics and Astronomy, University of Nevada, Las Vegas, 4505 S. Maryland Pkwy, Las Vegas, NV 89154-4002, USA
\and
Humanities and Sciences/Museum Careers, Musashino Art University, Tokyo 187-8505, Japan
\and
INAF – Osservatorio Astronomico di Trieste, I-34131 Trieste, Italy
\and 
Cosmic Dawn Center (DAWN), Copenhagen, Denmark
\and
DTU Space, Technical University of Denmark, DK-2800 Kgs. Lyngby, Denmark
}

\date{Received -- / Accepted --}

\abstract{Numerical efforts increasingly resolve the small-scale structure of the circumgalactic medium (CGM), but the dynamical impact of ionising radiation on its cold $10^4\, \rm K$ component remains poorly understood.} {We investigated the evolution of static cold gas structures exposed to the extreme-ultraviolet (EUV) radiation of quasars.} {We developed an analytical framework to describe the evolution of such clouds, introducing a new threshold that defines when a cloud becomes radiation-shielded. The framework was validated using radiation-hydrodynamic simulations of single static clouds.} {The framework predicts three evolutionary regimes: (i) an optically thin regime, in which radiation uniformly ionises the cloud; (ii) a radiation-shielded regime, where the cloud remains largely unaffected; and (iii) a "rocket-effect" regime, in which an ionisation front propagates through the cloud, ionising the illuminated side while compressing the opposite side and accelerating the surviving cold clump. In this latter regime, the cloud's Lyman-$\alpha$ (Ly$\alpha$) luminosity can increase by up to one order of magnitude compared to the optically thin case. Such luminosities are as high as $70\%$ of the values obtained from a fluorescent regime without considering hydrodynamical response. Unless the cloud is self-shielded, at least $\sim 50-60\,\%$ of Ly$\alpha$ emission arises from recombination. Applying this framework to both a population of clouds along a line of sight and a ray propagating inside a single cold stream, we find that the cold inter-galactic medium (IGM) and CGM around bright quasars ($L_{\mathrm{\nu,LL}} \sim 10^{31.6} \, \rm erg\, s^{-1}\, Hz^{-1}$) is likely fully ionised, whereas the one around faint quasars ($L_{\mathrm{\nu,LL}} \sim 10^{28.6} \, \rm erg\, s^{-1}\, Hz^{-1}$) predominantly experiences a rocket-effect regime.}{These results imply that the hydrodynamical response of cold CGM structures to quasar radiation must be considered when deriving CGM and IGM physical properties from observations, particularly for faint quasars.}

\keywords{Galaxies: evolution -- (Galaxies:) intergalactic medium -- Methods: numerical -- Methods: analytical -- (Galaxies:) quasars: general -- Radiation: dynamics}
\maketitle

\section{Introduction}
Cosmological zoom-in simulations are among the most realistic tools for predicting the state of gas around galaxies \citep[see][for reviews]{Faucher2023,Gronke2026}. However, recent cosmological galaxy-formation simulations show that key properties of the cold ($\sim 10^4\,\rm K$) gas, such as the total gas mass or the amount of cold clumps, do not numerically converge \citep[e.g.][]{Nelson2020,Ramesh2024}.
In other words, increasing the resolution continues to reveal ever smaller cold structures.
Moreover, high-resolution idealised simulations indicate that radiative cooling \citep{Mandelker2020a}, in addition to magnetic fields \citep[e.g.][]{Ledos2024b,Hidalgo2024,Kaul2025}, gravitational acceleration \citep{Aung2024}, and thermal conduction \citep{Sander2021,Bruggen2023,Ledos2024a}, can prolong the survival of cold gas structures.
Thus, the central question in CGM studies has gradually shifted: it is no longer merely whether cold gas can survive within hot haloes, but also how dense and how finely structured the cold CGM may be -- namely, its density distribution, the characteristic cold cloud sizes, and ultimately its impact on galaxy formation and evolution, particularly at high $z$ where the cold gas is denser and more abundant.
\\
Over the past decades, observations have shown that gas-surrounding galaxies on scales of hundreds of kiloparsec or more at z$>2$ can be directly detected in emission when illuminated by quasars \citep[e.g.][]{Cantalupo2014,Borisova16,Arrigoni2018,Fossati2021,Mackenzie2021}. In particular, integral field spectrographs such as the Multi Unit Spectroscopic Explorer (MUSE) \citep{Bacon2010} and the Keck Cosmic Web Imager (KCWI) \citep{Morrissey2018} can provide 3D maps of the CGM thanks to the availability of velocity information at every spatial location. Owing to its high emissivity, the hydrogen Lyman-$\alpha$ (Ly$\alpha$) is the most luminous emission line for the study of gas within the inter- and circumgalactic medium (IGM and CGM) detectable with current facilities. However, extracting physical properties such as gas density and kinematics from the observed Ly$\alpha$ emission is difficult, because of its multiple production mechanisms \citep{Cantalupo2014,Pezzulli2019,Ouchi2020}, each with different sensitivities to the gas thermodynamic state.
Therefore, the interpretation of the observed CGM Ly-alpha emission requires assumptions, which need to be tested with simulations. In this context, the impact of intense quasar radiation fields remains largely unexplored.
\\
The impact of radiation on cold gas has long been studied in other contexts, including the evaporation of mini-haloes by the ultraviolet background and the photo-evaporation of interstellar clouds exposed to stellar radiation.
In the latter case, numerous analytical studies \citep[e.g.][]{spitzer_physical_1978,bertoldi_photoevaporation_1989,bertoldi_photoevaporation_1990,gorti_photoevaporation_2002} have focused on the evolution of the ionised boundary layer and the shock compression of the cloud interior. These models were later refined through 1D and 2D simulations \citep[e.g.][]{lefloch_cometary_1994,mellema_photo-evaporation_1998,Iliev2006} and eventually extended to fully 3D studies \citep[e.g.][]{raga_photoevaporation_2005,Decataldo2019,Nakatani2019,Nakatani2020}.
\\
However, these studies typically focus on denser ($\gtrsim 10^3\,\rm cm^{-3}$), colder ($<10^2\,\rm K$) gas and do not probe the regime relevant to the IGM and the CGM in the vicinity of quasars.

In this work, we address this gap by developing an analytical framework, supported by radiation-hydrodynamic simulations, to investigate the evolution of cold IGM and CGM clouds exposed to quasar radiation.
\\
We first introduce and extend the analytical framework describing the evolution of individual cold clouds in Sect.~\ref{sec:theory}. Section~\ref{sec:num} details the numerical setup of our simulation suite. The validation of the analytical framework and the main simulation results are discussed in Sect.~\ref{sec:results}. In Sect.~\ref{sec:discussion}, we extend the single cloud model to an ensemble of clouds, before concluding in Sect.~\ref{sec:ccl}.

\section{Analytical model for single cloud evolution}\label{sec:theory}
\begin{figure*}
    \centering
    \includegraphics[width=.99\linewidth]{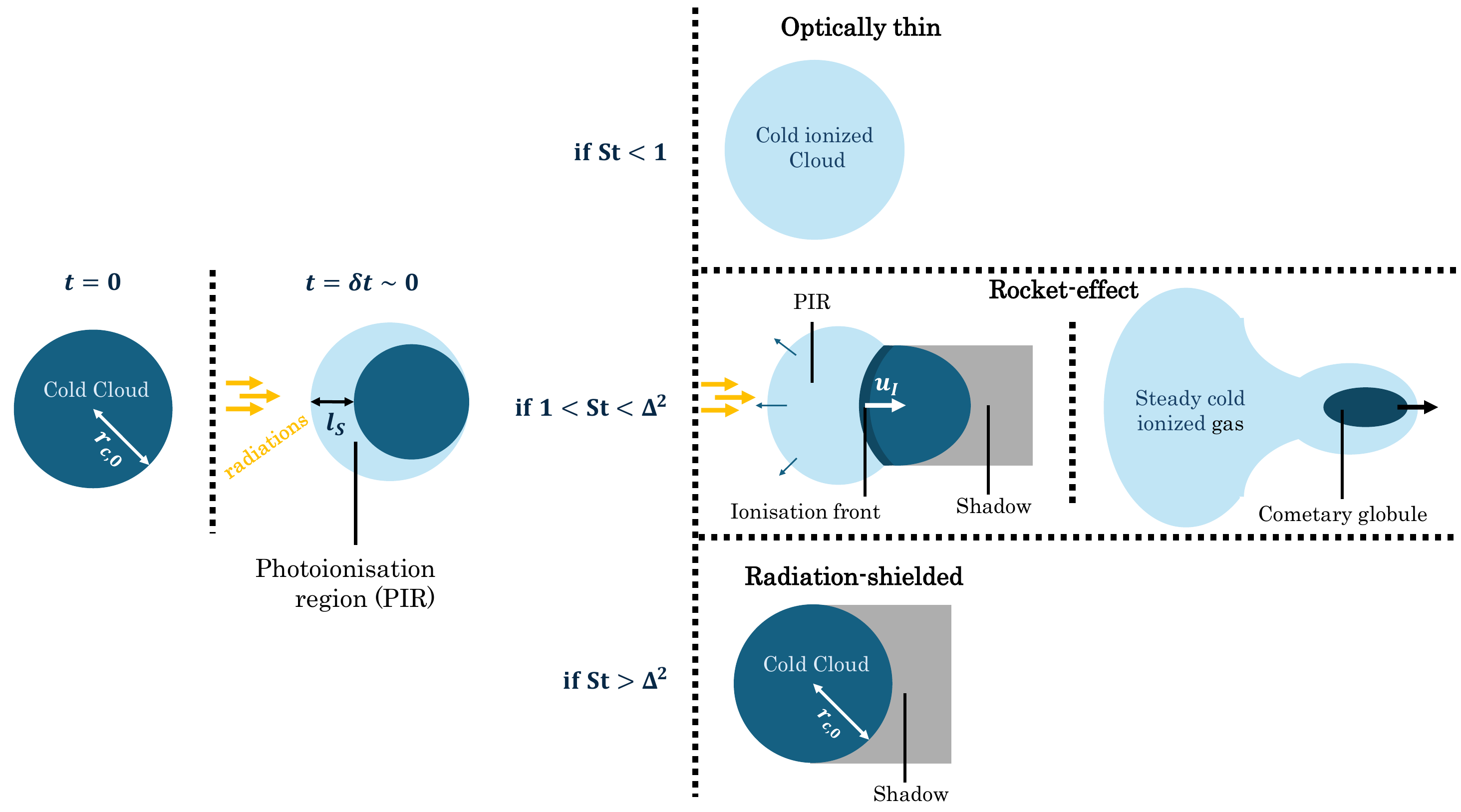}
    \caption{Illustration of the different evolutionary paths of the cloud illuminated by a quasar. Starting from an initial neutral cloud, which is suddenly illuminated by EUV radiation at an infinitesimal time $\delta t$. The Strömgren number $\mathrm{St}\propto r_{\mathrm{c,0}}n_{\mathrm{H,0}}^2F_{\mathrm{q}}^{-1}$ (Eq.~\ref{eq:St}) defines the cloud's evolution as described in Sect.~\ref{sec:theory}. The cloud follows three characteristic regimes: an optically thin regime, a rocket-effect regime, and a radiation-shielded regime.}
    \label{fig:model_illust}
\end{figure*}

\citet{bertoldi_photoevaporation_1989}, \citet{lefloch_cometary_1994}, and \citet{gorti_photoevaporation_2002} introduced a dimensionless parameter, later termed as the Strömgren number \citep[e.g.][]{Shapiro2004,Iliev2006}, that characterises the evolution of irradiated clouds.
Here, we present this classification framework and extend it, as illustrated in Fig.~\ref{fig:model_illust}, further differentiating the expected radiation-hydrodynamical responses of cold clouds to photoionisation.
As shown in the left panel of Fig.~\ref{fig:model_illust}, we consider a uniform cloud of radius $r_{\mathrm{c,0}}$, temperature $T_0$, and neutral hydrogen number density $n_{\mathrm{H,0}}$, illuminated on one side by a unidirectional flux of ionising photons $F_{\mathrm{q}}$ (in unit of photons per centimetre squared per second).
The primary mechanism that balances ionisation from quasar extreme UV (EUV) radiation is hydrogen recombination. Along a given ray, this balance can be written as
\begin{equation}\label{eq:l_S}
    l_{\mathrm{S}} \times n_{\mathrm{H,0}}^2 \alpha^{\mathrm{B}}_{\mathrm{H}} = F_{\mathrm{q}},
\end{equation}
where $l_{\mathrm{S}}$ is the Strömgren length, defined as the length over which the ionising photon flux is balanced by recombination. We used the recombination rate coefficient from \citet{Cen1992} $\alpha^{\mathrm{B}}_{\mathrm{H}}= \alpha^{\mathrm{B}}_{\mathrm{H}}(T_{\mathrm{i}})$ defined from the temperature of the ionised cold gas $T_{\mathrm{i}}$ defined later in this section. We adopted the on-the-spot-approximation (case~B)\footnote{ Using case~A approximation would only lead to a recombination coefficient higher by roughly a factor 1.6.}. The corresponding Strömgren number is hence defined as
\begin{equation}\label{eq:St}
\mathrm{St} = \frac{2r_{\mathrm{c,0}}}{l_{\mathrm{S}}} = \frac{2r_{\mathrm{c,0}}n_{\mathrm{H,0}}^2 \alpha^{\mathrm{B}}_{\mathrm{H}}}{F_{\mathrm{q}}}.
\end{equation}
This quantity can also be interpreted as the ratio between the cloud column density and the ionised column density \citep{bertoldi_photoevaporation_1989,Nakatani2019,Nakatani2020}.

As illustrated in Fig.~\ref{fig:model_illust}, when $\mathrm{St}<1$ or $2r_{\mathrm{c,0}}<l_{\mathrm{S}}$, the radiation can fully ionise the cloud in a timescale far below any other relevant one. The cloud is then uniformly heated and expands until reaching pressure equilibrium with the surrounding medium. We refer to this case as the optically thin regime.

When $\mathrm{St}>1$ or $2r_{\mathrm{c,0}}>l_{\mathrm{S}}$, radiation initially penetrates only a layer of thickness $l_{\mathrm{S}}$ within the cloud. The stability of this layer, hereafter referred to as the photoionisation region (PIR), governs the subsequent evolution of the cloud. We named this region analogously to the photodissociation regions (PDRs) of interstellar clouds \citep[e.g.][]{gorti_photoevaporation_2002} in which the radiation photodissociates \ion{H}{2} molecules.
\\
Compared to previous studies, we consider that, within the PIR, the ionised gas can also satisfy a balance between ionisation and recombination:
\begin{equation}\label{eq:l_S2}
    l_{\mathrm{S,i}} \times n_{\mathrm{H,i}}^2 \alpha^{\mathrm{B}}_{\mathrm{H}} = F_{\mathrm{q}},
\end{equation}
where $l_{\mathrm{S,i}}$ is the Strömgren length within the PIR and $n_{\mathrm{H,i}}$ is the ionised hydrogen number density, once pressure equilibrium with the ambient medium is established.
Analogously, we define for the PIR a Strömgren number,
\begin{equation}\label{eq:St_i}
    \frac{2r_{\mathrm{c,0}}}{l_{\mathrm{S,i}}} = \mathrm{St}\times\left(\frac{n_{\mathrm{H,i}}}{n_{\mathrm{H,0}}}\right)^2= \mathrm{St}\times\Delta^{-2},
\end{equation}
where
\begin{equation}\label{eq:Delta}
\Delta\equiv n_{\mathrm{H,0}} / n_{\mathrm{H,i}}= \mu_{c,0}T_{\mathrm{i}} / T_{\mathrm{c,0}}\mu_{c,i}\sim 2T_{\mathrm{i}} / T_{\mathrm{c,0}}
\end{equation}
is the density contrast between neutral and ionised gas, here approximated by the corresponding temperature ratio. We treated $\Delta$ as a constant by adopting a crude characteristic ionised-gas temperature $T_{\mathrm{i}}\sim10^{4.3}\,\rm K$, consistent with our simulations in the relevant region of the parameter space. In reality, however, $T_{\mathrm{i}}$ depends on both $n_{\mathrm{H,0}}$ and $F_{\mathrm{q}}$. For an initial cloud temperature $T_{\mathrm{c,0}}=6500\,\rm K$, this yields $\Delta^2 \sim 37$.
\\
The condition $\mathrm{St}/\Delta^2=1$, or equivalently $2r_{\mathrm{c,0}}=\Delta^2 l_{\mathrm{S}}$, marks the condition at which recombination balances photoionisation within the PIR.

If $\mathrm{St}>\Delta^{2}$ or $r_{\mathrm{c,0}}>\Delta^{2}l_{\mathrm{S}}$, the PIR cannot be fully sustained and remains limited within a depth of size much smaller than $l_{\mathrm{S}}$, while the ionisation front is stalled. The cloud is therefore shielded from radiation, defining the limiting case: the radiation-shielded regime.

If $1<\mathrm{St}<\Delta^{2}$ or $l_{\mathrm{S}}<r_{\mathrm{c,0}}<\Delta^{2}l_{\mathrm{S}}$, the PIR can develop and an ionisation front propagates into the clouds. As shown in the middle panel of Fig.~\ref{fig:model_illust}, the heating and expansion of the ionised gas compress the neutral material ahead of the front. Once the ionisation front has traversed the cloud, the back reaction of the expanding photoionised gas launches the remaining compressed neutral gas, whose structure is then termed "cometary globule" \citep[e.g.][]{spitzer_physical_1978,bertoldi_photoevaporation_1990,gorti_photoevaporation_2002,Nakatani2019}. This behaviour is akin to a rocket being propelled by its exhaust, thus giving the name of the "rocket-effect" regime. In this regime, and within our parameter range, the ionisation front velocity can be approximated as
\begin{equation}
    u_{\mathrm{I,c}} = F_{\mathrm{q}}/n_{\mathrm{H,0}}.
\end{equation}
A detailed description of ionisation front types, along with an equivalent derivation of our analytical framework based on $u_{\mathrm{I,c}}$, is provided in Appendix~\ref{app:theory}.
\\
In summary, the evolution of dense cold CGM clouds can be classified according to their Strömgren number $\mathrm{St}$ (Fig.~\ref{fig:model_illust}) into three distinct regimes: an optically thin regime ($\mathrm{St}<1$), a rocket-effect regime ($1<\mathrm{St}<\Delta^2$), and a radiation-shielded regime ($\mathrm{St}>\Delta^2$). 
While the distinction between the optically thin ($\mathrm{St}<1$) and partial-ionisation ($\mathrm{St}>1$) limits is well established in the literature, we identify for the first time, to our knowledge, a second critical threshold, $\mathrm{St}=\Delta^{2}$
, which separates the rocket-effect and radiation-shielded regimes\footnote{
The value $\Delta^2\propto T_{\mathrm{c,0}}^{-2}$ depends on the physical context and sets the extent of the PIR sustained regime.
In previous studies of interstellar clouds focusing on the rocket-effect \citep[e.g.][]{bertoldi_photoevaporation_1989,lefloch_cometary_1994,Iliev2009,Nakatani2019}, the characteristic cloud temperature $T_{\mathrm{c,0}}\sim 100\,\rm K$ implies $\Delta^2 \gtrsim 10^4$. Consequently, the rocket-effect regime spans a much broader range of densities, cloud radii, and incident photon fluxes, while the radiation-shielded regime is reached only for substantially weaker irradiation or larger cloud column densities. This regime therefore lies well outside the parameter space explored in those studies and could safely be neglected.}.

Finally, in our above framework, for fixed cloud radius and initial temperature, the cloud evolution depends on two physical parameters: the initial hydrogen number density $n_{\mathrm{H,0}}$ and the incident ionising photon flux $F_{\mathrm{q}}$. While $\mathrm{St}$ captures their combined effect on the cloud state, a second independent parameter is required to fully span the $(n_{\mathrm{H,0}}, F_{\mathrm{q}})$ space. We therefore constructed a complementary dimensionless variable, $\Upsilon$ (capital upsilon), defined such that it is orthogonal to $\mathrm{St}$ in parameter space:
\begin{equation}\label{eq:upsilon}
    \Upsilon \equiv An_{\mathrm{H,0}}^{a} F_{\mathrm{q}}^{2a},
\end{equation}
where $A=1\, \rm cm^{7/2}\, s$ and $a=0.5$ are arbitrary constants. As shown in Sect.~\ref{sec:results}, $\Upsilon$ traces the strength of photoionisation, while $\mathrm{St}$ governs the dynamical response of the cloud. Our choice of $\Upsilon$ over the dimensionless ionisation parameter $\mathcal{U}= F_{\mathrm{q}}/cn_{\mathrm{H,0}}$ was motivated by the fact that $(\mathcal{U},\mathrm{St})$ are not orthogonal coordinates. We, however, present results in terms of $\mathcal{U}$ in Appendix~\ref{app:fits4obs}.
\\
The evolution of an irradiated cloud is thus fully characterised by the pair $(\mathrm{St}, \Upsilon)$, which respectively describe the cloud response and the strength of the ionising field.

\section{Numerical simulations}\label{sec:num}
We validated the analytical framework through a suite of radiation-hydrodynamic simulations. In this section, we outline the relevant parameter space, adopted initial conditions, and computational method.

\subsection{Expected densities and radiation field}\label{sec:sub:param}
The CGM of massive haloes is expected to contain diffuse hot gas reaching temperatures of $T\gtrsim10^6\,\rm K$ through virial shock-heating, together with colder and denser relatively pristine gas accreting along intergalactic filaments with temperatures of $T\lesssim10^4\,\rm K$ \citep[see review by ][]{Tumlinson2017}. The temperature of this cold component is regulated by the balance between radiative cooling and photo-heating from the ambient UV background, while its commonly low metallicity limits additional metal-line cooling. Such multiphase environments are extensively observed through Ly$\alpha$ emission and absorption and are well studied within cosmological zoom-in simulations \citep[e.g.][]{Nelson2020,Ramesh2024, Waterval2025}. Building upon these earlier simulations, \citet{Dekel2013} introduced a toy model for the cold gas accretion in massive haloes ($M_{\rm h}>10^{11},\rm M_\odot$) at $z>1$, later refined by \citet{Mandelker2020b} and \citet{Aung2024}.
These models predict cold-stream densities of $\sim 1.5\times10^{-3}\,\mathrm{cm^{-3}}$ near the virial radius of a $10^{12}\,\mathrm{M_\odot}$ halo at $z\sim2$, rising up to $\lesssim 300\,\mathrm{cm^{-3}}$ at $\sim 10\,\mathrm{pkpc}$ in more massive haloes ($10^{13}\,\mathrm{M_\odot}$) at $z\sim3.5$. The corresponding temperature contrast, $\delta = T_{\mathrm{hot}}/T_{\mathrm{c,0}}$, can reach values of $\sim 2000$.

The radiation field considered in this work is motivated by bright EUV sources, in particular quasars. We focus on luminosities typical of radio-quiet QSOs at $z>3$ \citep{Cantalupo2014,Borisova16,Arrigoni2018,Fossati2021,Mackenzie2021}. 
The specific luminosity at the Lyman limit spans $L_{\nu,\mathrm{LL}} \sim 10^{28.6}$ to $10^{31.6}\,\mathrm{erg\,s^{-1}\,Hz^{-1}}$. Their spectral energy distribution between $1$-$100\,\mathrm{Ryd}$ is commonly described by a power law $L_\nu \propto \nu^{-p}$ with $p \simeq 1.7$ \citep{Lusso2015}, corresponding to ionising photon production rates up to $\dot{N}_{\mathrm{ph}} \lesssim 10^{58}\,\mathrm{s^{-1}}$. 
The relevant quantity for cloud evolution is the incident photon flux,
\begin{equation}
    F_{\mathrm{q}}=\frac{\dot{N}_{\mathrm{ph}}}{4\pi r^2},
\end{equation}
where $r$ is the distance from the quasar. For clouds located at $10$-$100\,\mathrm{pkpc}$, this corresponds to fluxes up to $\sim10^{10}$-$10^{12}\,\mathrm{cm^{-2}\,s^{-1}}$.

\subsection{Initial conditions} \label{sec:sim_description}
The simulations were designed to probe the transitions between the regimes identified in Sect.~\ref{sec:theory}, with particular emphasis on the onset of the rocket-effect regime.
\begin{figure}
    \centering
    \includegraphics[width=.99\linewidth]{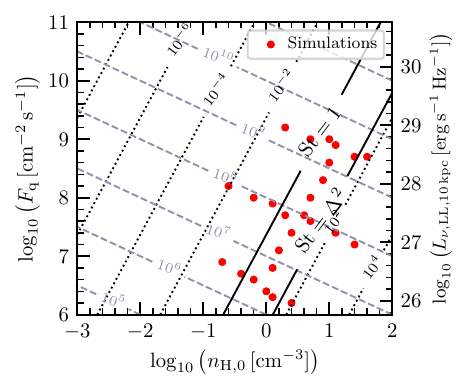}
    \caption{$\mathrm{St}$ (dotted black) and $\Upsilon$ (dashed grey) parameter contour lines for typical values of $n_{\mathrm{H,0}}$ and $F_{\mathrm{q}}$ in the CGM and a cloud radius of $50\, \mathrm{pc}$. The left y-axis shows the corresponding quasar Lyman-limit specific luminosities $L_{\mathrm{\nu,LL}}$ assuming a cloud at a distance of $10 \, \rm pkpc$ from the quasar. The red dots represent the simulation's initial parameters.}
    \label{fig:nH_Fq_eta_map}
\end{figure}
Figure~\ref{fig:nH_Fq_eta_map} shows the location of the simulations in the $(n_{\mathrm{H,0}},F_{\mathrm{q}})$ plane. We adopt a cloud radius $r_{\mathrm{c,0}}=50\,\mathrm{pc}$, representative of small, dense CGM clumps \citep{Nelson2020,Yao2025}, and express the photon flux in terms of the corresponding $L_{\nu,\mathrm{LL}}$ at a distance of $10\,\mathrm{pkpc}$.
We focus on relatively low photon fluxes ($F_\mathrm{q} \lesssim 10^9 \,\rm cm^{-2}\, s^{-1}$), as this range allows all three regimes described in Sect.~\ref{sec:theory} to manifest over densities relevant for the cold CGM (Fig.~\ref{fig:nH_Fq_eta_map}).

We performed six, eight, and six simulations at fixed $\Upsilon \simeq 10^{6.5}$, $10^{8}$, and $10^{9.5}$, respectively, thereby sampling different ionisation strengths across a range of $\mathrm{St}$.
An additional six simulations at fixed $\mathrm{St} \sim 8$ were used to isolate the effect of $\Upsilon$ within the rocket-effect regime. 
We adopted approximately evenly spaced steps in ($\mathrm{St}$ and $\Upsilon$), while including small random perturbations, visible as small offsets in Fig.~\ref{fig:nH_Fq_eta_map}.

The computational domain is a cubic box of size $L=12.8\,r_{\mathrm{c,0}}$ of hot CGM gas. At its centre lies a static spherical cloud with fixed temperature $T_{\mathrm{c,0}}\simeq 6500\,\mathrm{K}$. This fiducial cloud temperature was chosen to represent partially photo-heated cold neutral gas near the thermal equilibrium temperature expected under UV background radiation. The surrounding hot medium has a temperature $T_{\mathrm{CGM}}=T_{\mathrm{c,0}}\delta$. For dense clouds, we adopted $\delta=1000$ to effectively suppress cooling of the background gas over the simulation time. For lower-density cases ($n_{\mathrm{H,0}} < 0.5\,\mathrm{cm^{-3}}$), we used $\delta = 400$.

The base adaptive mesh refinement (AMR) level leads to the resolution of $\Delta x=L/2^5$ per dimension. We define the highest resolution level following a 'verification' procedure using grid convergence study \citep{Roache1998}, described in Appendix~\ref{app:GC}. In practice, this leads to maximum AMR levels of 8 and 9, giving 40 and 80 cells per cloud's diameter, respectively.

\subsection{Governing equations and computational methods}\label{sec:gov_eq} 
We solved the radiation-hydrodynamics equations in conservative form using the code RAMSES-RT \citep{Teyssier2002,Rosdhal2013}, as
{\small
\begin{equation}\label{eq:RHD_system_1}
\left\{ \def\arraystretch{0.5}\begin{array}{l}
\dfrac{\partial\rho}{\partial t}+\nabla\cdot\left(\rho\mathbf{u}\right)=0\,,\\
\\
\dfrac{\partial\left(\rho\mathbf{u}\right)}{\partial t}+\nabla\cdot\left[\rho\left(\mathbf{u}\otimes\mathbf{u}\right) + p\mathbb{I}\right] =0\,,\\
\\
\dfrac{\partial e}{\partial t}+\nabla\cdot\left[\left(e+p\right)\mathbf{u}\right] = \Lambda\left(\rho,e_{\mathrm{in}}\right) \,,\\
\\
\dfrac{\partial N_\nu}{\partial t}+\nabla\cdot\mathbf{F}_\nu = -cN_\nu\left(\mathbf{n}\cdot\mathbf{\sigma}_\nu^{N}\right)  \,,\\
\\
\dfrac{\partial \mathbf{F}_\nu}{\partial t}+c^2\nabla\cdot\mathbb{P}_\nu = -c\left(\mathbf{n}\cdot\mathbf{\sigma}_\nu^{N}\right)\mathbf{F}_\nu  \,,
\end{array}\right.
\end{equation} }
where $\rho$, $\mathbf{u}$, $p$,  $e$ denote the gas density, velocity, pressure, and total energy density, while $N_\nu$ and $\mathbf{F}_\nu$ are the photon number density and flux in each photon group. $\mathbb{I}$ is the identity matrix. The total energy density is given by
\begin{equation} \label{eq:RHD_system_2}
e = e_{\mathrm{in}} + \rho\frac{u^2}{2} =\frac{p}{\gamma - 1} + \rho\frac{u^2}{2},
\end{equation}
where $\gamma$ is the adiabatic index. The radiation pressure tensor $\mathbb{P}_\nu$ was computed using the Eddington tensor approximation (M1-closure), which is appropriate in the present configuration, as we considered a single radiation source and adopted the Case B approximation \citep{Rosdhal2013,Rosdahl2015}. 

RAMSES-RT employs the reduced speed-of-light approximation, for which we adopted $c_{\mathrm{r}} = 0.01\,c$. The validity of this approximation requires three conditions \citep{Rosdhal2013,Deparis2019,Ocvirk2019}. First, the light crossing time of the cloud $t_{\mathrm{light}}=2r_{\mathrm{c,0}}/c$ must be much shorter than the recombination time $t_{\mathrm{rec}}$. Second, the simulation time, here $t_{\mathrm{sim}}=20\,\rm Myr$, must exceed $t_{\mathrm{rec}}$. Third, the ionisation front velocity $u_{\mathrm{I,c}}$ must remain below $c_{\mathrm{r}}$. In our simulations, these conditions were satisfied with $t_{\mathrm{light}}/t_{\mathrm{rec}} \in [10^{-4},10^{-2}]$, $t_{\mathrm{rec}}/t_{\mathrm{sim}} \in [10^{-4},10^{-1}]$, and $u_{\mathrm{I,c}} \lesssim c_{\mathrm{r}}$\footnote{In one simulation with the highest $F_{\mathrm{q}}$, we find $u_{\mathrm{I,c}} \sim 2.6c_{\mathrm{r}}$. However, this case lies in the optically thin regime, where no ionisation front is formed, and the reduced speed of light does not affect our conclusions.}.

Radiation was discretised into $M=6$ photon groups\footnote{Our choice of $M=6$ was derived from a convergence study for the number of photon groups detailed in Appendix~\ref{app:phot_conv}.} spanning over the ionisation thresholds of \ion{H}{I}, \ion{He}{I}, and \ion{He}{II}. The gas chemistry includes hydrogen and helium with absorbing species $\mathbf{n}=(n_{\mathrm{H_I}},n_{\mathrm{He_I}},n_{\mathrm{He_{II}}})$. Photon-group cross sections are frequency-averaged using the assumed source spectrum,
\begin{equation} \label{eq:RHD_system_3}
\mathbf{\sigma}_{\mathrm{X},\nu}^{N} = \frac{\int_{\nu_0}^{\nu_1} \sigma_{\mathrm{X}}\left(\nu\right)J\left(\nu\right) /h\nu \,\mathrm{d}\nu }{\int_{\nu_0}^{\nu_1} J\left(\nu\right) /h\nu \,\mathrm{d}\nu },
\end{equation}
where $J(\nu)$ follows the adopted spectral energy distribution (SED). These cross sections were precomputed, assuming a time-independent radiation source. Atomic cross sections were taken from \citet{Verner1996}. The non-equilibrium chemistry network was solved implicitly at each time step to compute the net heating and cooling term $\Lambda$.

The hydrodynamics was solved using the HLLC Riemann solver \citep{Toro1994} with a second-order MUSCL-Hancock Godunov scheme \citep{vanLeer1985}. Radiation transport uses the HLL flux function \citep{Harten1983} to accurately capture the cloud shadow\citep{Rosdhal2013,Rosdahl2015}.

\section{Results}\label{sec:results}
We first examine the evolution of clouds across the regimes defined in Sect.~\ref{sec:theory}. We then turn to the resulting density fluctuations and their associated Ly$\alpha$ and H$\alpha$ emission.

\subsection{Evolution of the clouds}
\begin{figure*}
    \centering
    \includegraphics[width=.9\linewidth]{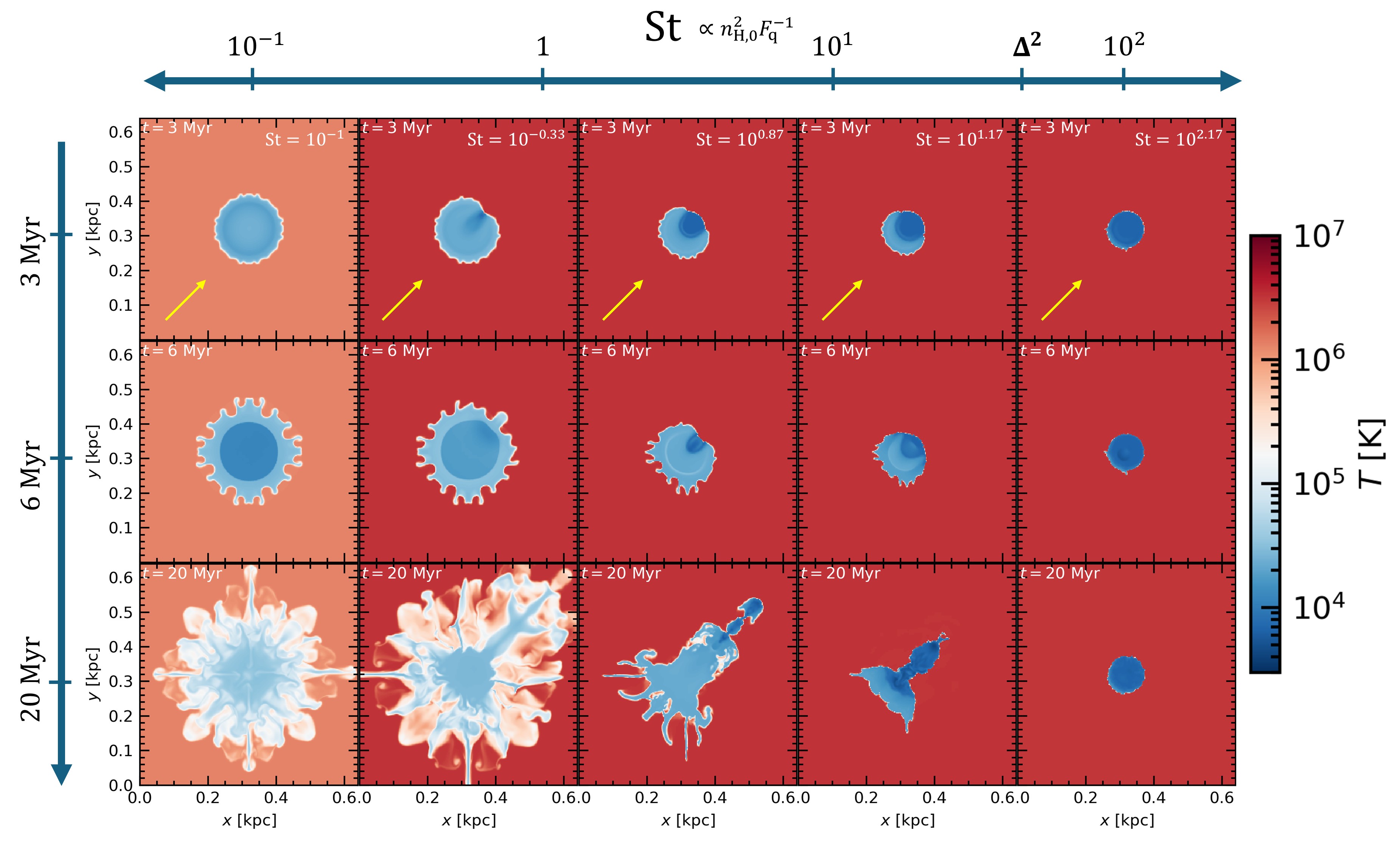}
    \caption{Temperature colour map slices, showing cloud evolution at $t = 3, 6,$ and $20\,\mathrm{Myr}$ for five representative simulations with different values of $\mathrm{St}$ (Eq.~\ref{eq:St}) and the ratio of the recombination time over the ionisation front crossing time in the cold cloud at fixed $\Upsilon \sim 10^8$. In all maps, quasar radiation enters from the bottom-left corner as shown by the yellow arrow.}
    \label{fig:evo_cloud_eta}
\end{figure*}
We begin with a qualitative overview of cloud evolution. Fig.~\ref{fig:evo_cloud_eta} shows temperature maps at $t = 3, 6,$ and $20\,\mathrm{Myr}$ for varying $\mathrm{St}$ at fixed $\Upsilon \sim 10^8$. Radiation enters from the bottom-left, as indicated by the yellow arrow. The simulations broadly follow the behaviour predicted in Sect.~\ref{sec:theory}, indicating that $\mathrm{St}$ is a robust predictor of the cloud’s evolutionary regime. For $\mathrm{St} < 1$ (optically thin regime), the cloud is uniformly heated and expands nearly isotropically into the surrounding medium. This expansion induces transient adiabatic cooling at early times, leading to a brief contraction before the cloud settles into pressure equilibrium at $t \gtrsim  5-10\,\mathrm{Myr}$. Owing to the sharp density contrast, numerical dispersion errors can seed Rayleigh-Taylor-like instabilities \citep[see discussion of the carbuncle artefact in][]{Iliev2009}, giving rise to a mixing layer between the cold and hot gas\footnote{Such instabilities may also have a physical origin, as suggested by \citet{Proga2015} and \citet{Waters2019}. However, our set-up differs from these studies, making it difficult to assess their relevance in the present case.}.
For $\mathrm{St} > \Delta^2$ (the radiation-shielded regime), the cloud remains largely unaltered, experiencing only minor perturbations.
In the rocket-effect regime ($\mathrm{St} \in [1,\Delta^2]$), the cloud evolves as shown in Fig.~\ref{fig:model_illust}. At the base of the ionisation front, the PIR expansion compresses the neutral gas side. Once the remnant neutral gas, the cometary globule, cannot be further compressed, it is expelled by the back-reaction of the PIR - the rocket effect. As in the optically thin regime, instabilities developing at the surface of the PIR give rise to finger-like structures.
\begin{figure*}
    \centering
    \includegraphics[width=.9\linewidth]{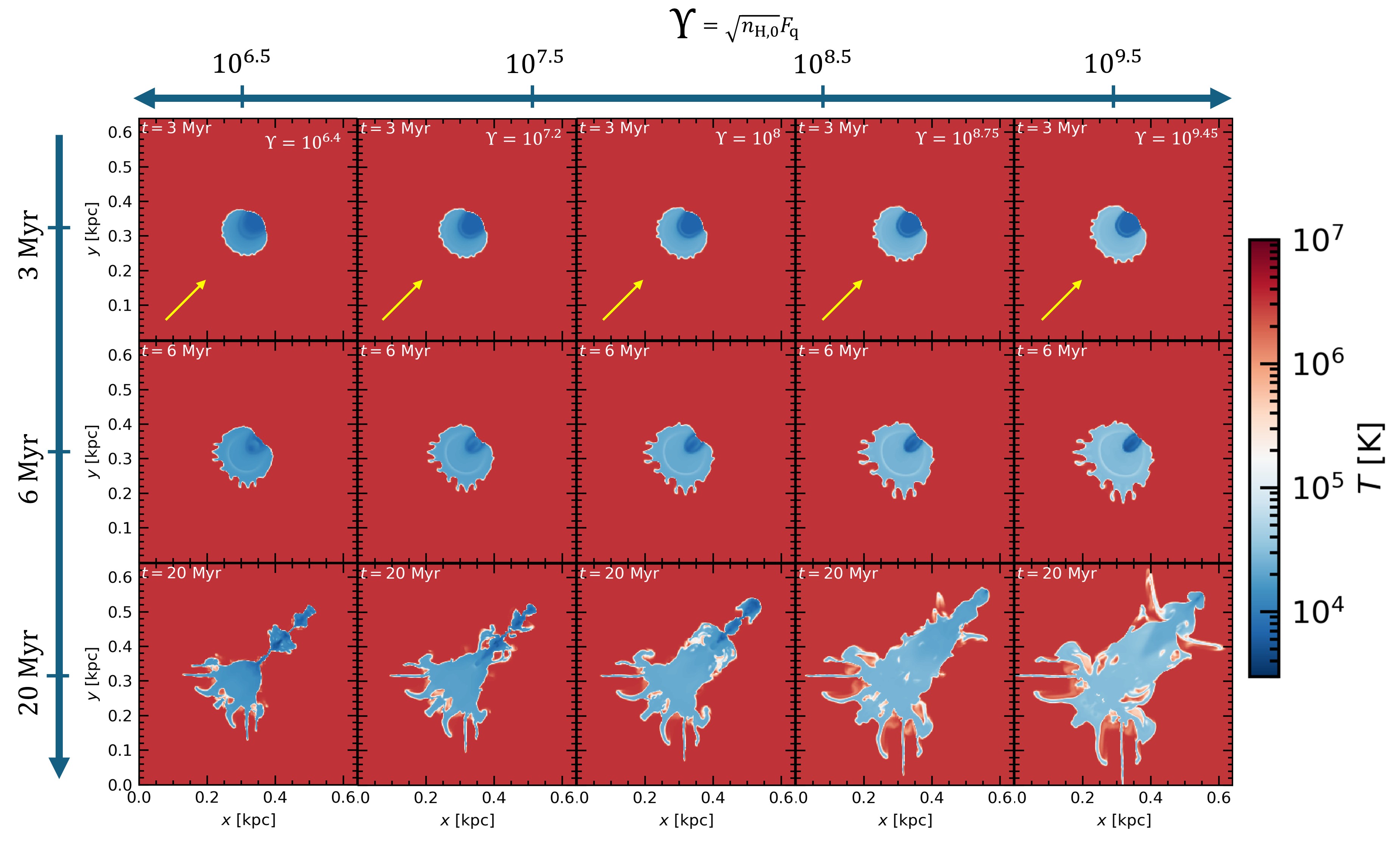}
    \caption{Same as Fig.~\ref{fig:evo_cloud_eta} but for different values of the ionisation strength parameter $\Upsilon$ at fixed $\mathrm{St}\sim 8$ (rocket-effect regime).}
    \label{fig:evo_cloud_ups}
\end{figure*}
\\
Figure~\ref{fig:evo_cloud_ups} illustrates the effect of the ionisation strength parameter $\Upsilon$ at fixed $\mathrm{St} \sim 8$ (rocket-effect regime). At all times, ionisation fronts reach similar positions, yielding broadly comparable clouds' morphologies. Increasing $\Upsilon$ primarily raises the temperature within the PIR and enhances the contrast with the dense, cold cometary globule, as the ionisation front velocity $u_{\mathrm{I,c}}$ becomes larger relative to the cloud sound speed.
Initially, ionisation fronts are R-type in all cases except for $\Upsilon \sim 10^{6.5}$, which exhibits an M-type front. As the PIR develops, it absorbs part of the photon flux, slowing the front and reducing its ability to ionise the remaining gas. Two competing effects then emerge: at low $\Upsilon$, the absorption of photons due to denser and cooler PIR slows and broadens the ionisation front; at high $\Upsilon$, photons efficiently cross the PIR, leading to strong compression, a denser globule and a hotter and more diffuse PIR.
\begin{figure*}
    \centering
    \includegraphics[width=.49\linewidth]{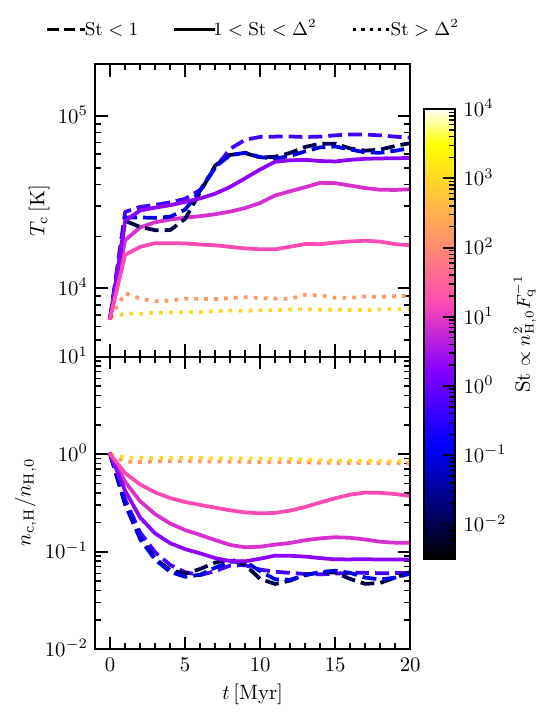}
    \includegraphics[width=.49\linewidth]{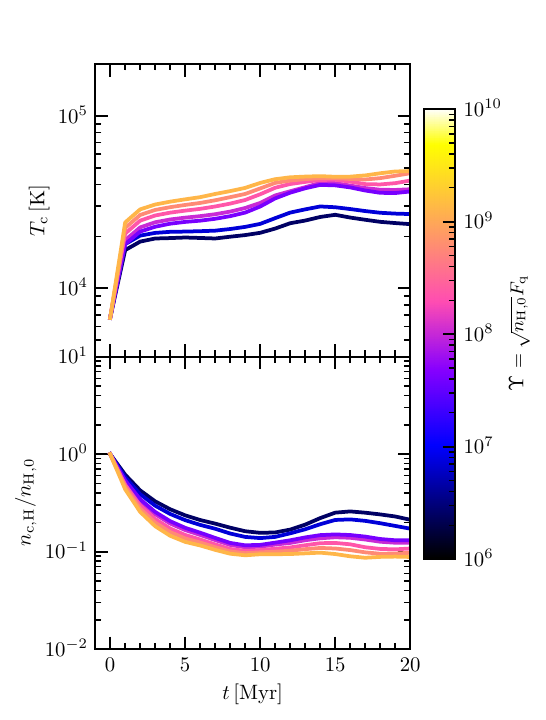}
    \caption{Cold gas mean temperature (top) and number density evolution (bottom). {\it Left panels:} Simulations at a fixed $\Upsilon\sim 10^8$ (equation~\ref{eq:upsilon}) and various $\mathrm{St}$ (Eq.~\ref{eq:St}). Simulations with clouds in the optically thin, radiation-shielded, and rocket-effect regimes are represented by the dashed, dotted and plain lines, respectively. {\it Right panels:} Simulations at a fixed $\mathrm{St} \sim 8$ (rocket-effect regime) with a varying ionisation strength parameter $\Upsilon$.}
    \label{fig:t_evo_cloud}
\end{figure*}
\\
Figure~\ref{fig:t_evo_cloud} shows the temporal evolution of the mean cold-gas temperature and hydrogen number density\footnote{The cold-gas temperature threshold follows the definition commonly used in studies of turbulent radiative mixing layers, $T_{\mathrm{thr}} = \sqrt{T_{\mathrm{c,0}}T_{\mathrm{CGM}}}$ \citep[see Appendix C of][]{Ledos2024a}.}. The corresponding thresholds are $\sim 10^{5.3}$ and $10^{5.1}\,\mathrm{K}$ for $T_{\mathrm{CGM}} \sim 10^{6.8}$ and $10^{6.4}\,\mathrm{K}$, respectively. In the left panel (fixed $\Upsilon \sim 10^8$), all simulations reach a quasi-steady state after a short transient phase ($t \sim 5$-$10\,\mathrm{Myr}$). The parameter $\mathrm{St}$ clearly governs the evolution: lower values correspond to stronger radiative impact, leading to hotter and more diffuse clouds. Notably, both the optically thin ($\mathrm{St} < 1$) and radiation-shielded ($\mathrm{St} > \Delta^2$) regimes showcase nearly constant temperatures and densities at fixed $\Upsilon$, suggesting that $\Upsilon$ is a good tracer of the characteristic thermodynamic states. For $\mathrm{St} < 1$, the clouds' temperature converges towards $T_{\mathrm{c}} \sim 7 \times 10^{4}\,\mathrm{K}$ and $n_{\mathrm{c,H}} \sim 0.06\,n_{\mathrm{H,0}}$, whereas for $\mathrm{St} > \Delta^{2}$ it remains cooler and denser, with $T_{\mathrm{c}} \sim 9 \times 10^{3}\,\mathrm{K}$ and $n_{\mathrm{c,H}} \sim 0.8\,n_{\mathrm{H,0}}$.
The right panel (fixed $\mathrm{St} \sim 8$) confirms that $\Upsilon$ primarily modulates the thermodynamic response within a given regime: higher values lead to progressively hotter and more diffuse clouds, without altering the qualitative evolutionary behaviour.

\subsection{Density fluctuations}
Our simulations resolve the dynamics of cold clumps that are typically unresolved in cosmological CGM simulations. To account for such subgrid density variations, the clumping factor $\mathcal{C}$ is commonly introduced to model their impact on hydrogen line emission \citep{Cantalupo2005,Cantalupo2014}. We therefore examined both the hydrogen number density probability distribution function (PDF) and the resulting clumping factors.

\subsubsection{Hydrogen number density PDF}\label{sec:pdf}
\begin{figure}
    \centering
    \includegraphics[width=1.\linewidth]{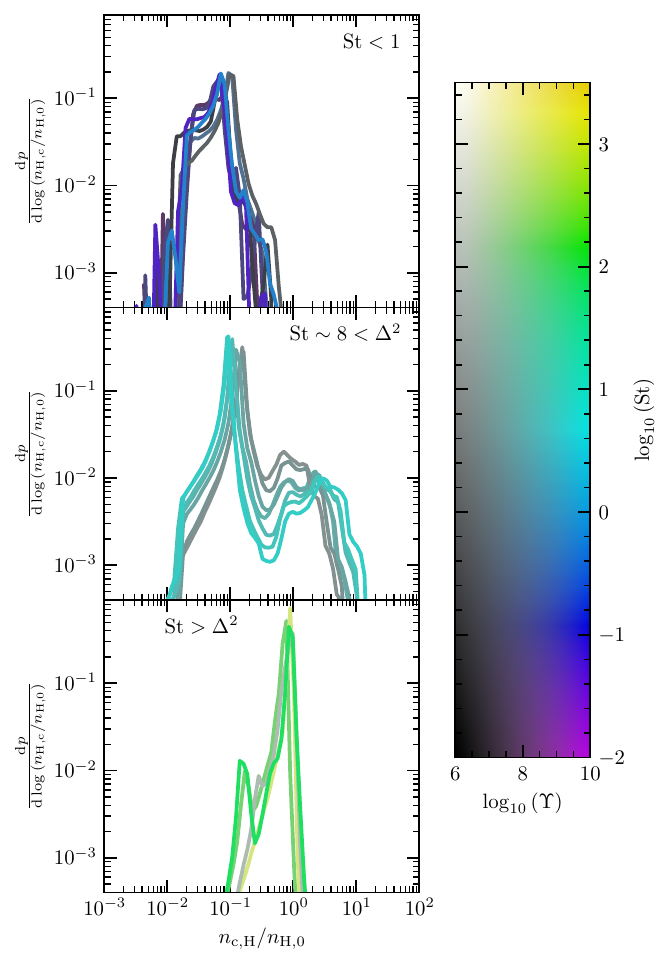}
    \caption{Cold-gas hydrogen number density PDFs weighted by density, normalised by $n_{\rm{H,0}}$ and averaged over $t\in [5,20]\,\rm Myr$. Top, middle, and bottom panels: Simulations in the optically thin, rocket-effect, and radiation-shielded regimes, respectively. Each line represents one simulation defined by a unique pair of $\left(\mathrm{St},\Upsilon\right)$. The parameters $\mathrm{St}$ and $\Upsilon$ are represented in the 2D colour map through colour variation and enforced white or black saturation, respectively. All simulations are shown except in the rocket-effect regimes, where, for clarity, only simulations with $\Upsilon\sim 10^8$ are plotted.}
    \label{fig:nH_pdf}
\end{figure}

We quantified the density fluctuations induced by the ionisation front by computing the PDF of the hydrogen number density. Figure~\ref{fig:nH_pdf} shows the cold-gas hydrogen number density PDFs, averaged over $t\in[5,20]\,\rm Myr$. The PDFs are weighted by the normalised densities $n_{\rm{H,c}}/ n_{\rm{H,0}}$, thereby emphasising the contribution of denser gas\footnote{The weighted PDFs can simply be expressed as $n\mathrm{d}p/\mathrm{d}n = \mathrm{d}p/\mathrm{d}\log(n)$.}.
The PDFs exhibit distinct patterns across the three regimes, consistent with the analytical boundaries $\mathrm{St} \in [1,\Delta^2]$ and validating further the framework introduced in Sect.~\ref{sec:theory}.
\\
In the optically thin regime ($\mathrm{St} < 1$), the density peak lies around $\sim 0.06\,n_{\mathrm{H,0}}$, as most of the cloud gas is ionised, settles into thermal equilibrium with the quasar radiation, and restores pressure equilibrium with the ambient gas.
\\
In the rocket-effect regime ($1<\mathrm{St}<\Delta^2$), the PDFs become bimodal, with peaks at $\sim 0.1\,n_{\mathrm{H,0}}$ and $\gtrsim n_{\mathrm{H,0}}$. The low-density peak traces ionised gas within the PIR, while the high-density component corresponds to the neutral gas and compressed cometary globule. Lower $\mathrm{St}$ and higher $\Upsilon$ reduce the amount of intermediate-density gas and shift the peaks towards more extreme values. This reflects the action of the ionisation front as rapidly propagating fronts ionise gas efficiently while inducing strong compression (Eq.~\ref{eq:jump_dens}), whereas slower fronts allow a higher amount of intermediate densities to persist.
\\
In the radiation-shielded regime ($\mathrm{St}>\Delta^2$), the PDF is dominated by neutral gas that largely retains its initial density. Nevertheless, a small peak appears near $\sim 0.2\,n_{\mathrm{H,0}}$, arising from perturbations that partially ionise the gas.

\begin{figure}
    \centering
    \includegraphics[width=.7\linewidth]{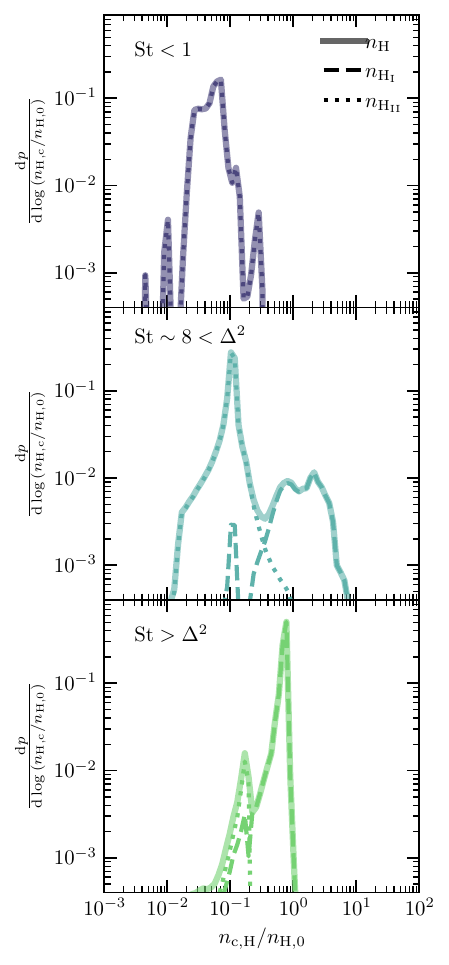}
    \caption{Cold-gas hydrogen number density PDFs weighted by number density and normalised to $n_{\rm H,0}$, averaged over $t\in[5,20]\,\rm Myr$. The plain, dashed, and dotted lines show total, neutral, and ionised hydrogen, respectively. Top, middle, and bottom panels: Representative simulations in the optically thin, rocket-effect, and radiation-shielded regimes. The colours match those in Fig.~\ref{fig:nH_pdf}.}
    \label{fig:nH_pdf_comp}
\end{figure}
Figure~\ref{fig:nH_pdf_comp} shows the density-weighted PDF decomposed into total, neutral, and ionised hydrogen for a representative simulation in each regime, all with $\Upsilon\sim 10^{8}$. This decomposition confirms that, across all regimes, the low- and high-density peaks are primarily associated with ionised and neutral gas, respectively.

\subsubsection{Hydrogen clumping factor}
The clumping factor directly reflects the width and asymmetry of the density PDF shaped by the ionisation front. For a given species $X$, it is defined as
{\small
\begin{equation} \label{eq:C_sim}
\mathcal{C}_{\mathrm{X,sim}} =   \left\langle n_{\mathrm{X}}^2\right\rangle_V \times \left\langle n_{\mathrm{X}}\right\rangle_V^{-2},
\end{equation}}
where the brackets $\left\langle -\right\rangle_V$ indicate the volume average of a given quantity over all the cold gas volume.
\begin{figure}
    \centering
    \includegraphics[width=.99\linewidth]{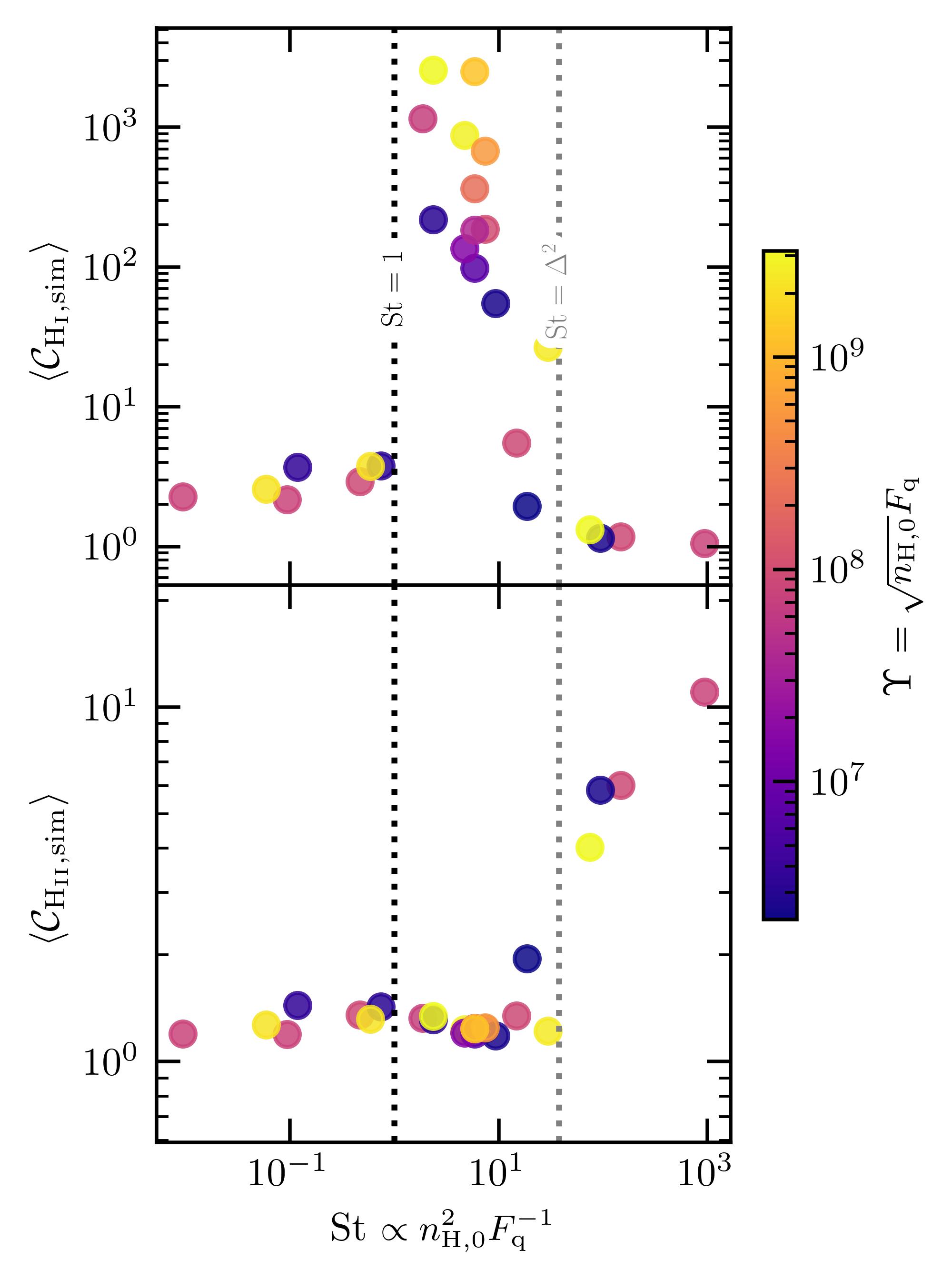} 
    \caption{Resulting cold gas clumping factor due to the effect of quasar radiation of the cold hydrogen gas, averaged over $t\in [5,20]\,\rm Myr$ for all simulations. Each circle represents a simulation positioned according to its parameters $\mathrm{St}$ (x-axis) and $\Upsilon$ (colour map). {\it Top panel:} Clumping factor of cold \ion{H}{I} gas. {\it Bottom panel:} Clumping factor of cold \ion{H}{II} gas.}
    \label{fig:clumping_f}
\end{figure}
Figure~\ref{fig:clumping_f} shows the time-averaged clumping factors of \ion{H}{I} and \ion{H}{II} over $t \in [5,20]\,\mathrm{Myr}$ for all simulations. Two runs, with $(\mathrm{St},\Upsilon)\sim (0.6,10^{9.3})$ and $(7.4,10^{8.8})$, do not fully converge in their \ion{H}{I} clumping factor; however, as discussed in Appendix~\ref{app:GC}, this has negligible impact on the derived luminosities and overall trends.

We emphasise that the clumping arises entirely from quasar irradiation, as the clouds are initially uniform with $\mathcal{C}=1$. For \ion{H}{I} gas, the clumping factor spans a wide range, from $\sim 2$ to $3000$ (median $\sim100$), but only in the rocket-effect regime, while remaining close to unity otherwise. In this regime, larger $\Upsilon$ and lower $\mathrm{St}$ lead to stronger compression and hence higher \ion{H}{I} clumping factors.
\\
As compression primarily affects the cold neutral gas ahead of the ionisation front, the \ion{H}{II} clumping factor remains close to unity. In the radiation-shielded regime, larger values of $\mathcal{C}_{\mathrm{H_{II},sim}}$ occur when the ionised gas constitutes only a small fraction of the cloud. For instance, $\mathcal{C}_{\mathrm{H_{II},sim}}\gtrsim 10$ arises from small ionised perturbations embedded within an otherwise neutral cloud, representing only a few per cent of the total cold hydrogen mass and volume (Fig.~\ref{fig:nH_pdf_comp}).

\subsection{Hydrogen lines}\label{sec:sub:emission}
A key challenge in both observations and radiative-transfer post-processing is distinguishing collisional excitation from recombination, as these processes respond differently to the number densities of hydrogen species. We derived polynomial fits to the H$\alpha$ and Ly$\alpha$ effective recombination rates $\alpha_{\mathrm{eff}}$, based on tabulated rates from \cite{Pengelly1964,Martin1988}. These fits are valid for $T \in [80,10^5],\rm K$ and achieve a maximum relative error of $0.25\%$ under case~B conditions.
Effective collisional excitation rates $\beta_{\mathrm{eff}}$ were computed following \cite{Giovanardi1987,Giovanardi1989}, also assuming case~B conditions. Further details on the effective rates and our fits are provided in Appendix~\ref{app:eff_rates}.
\\
The H$\alpha$ and Ly$\alpha$ luminosities were computed as
\begin{equation} \label{eq:L_rec}
\mathcal{L}_{\mathrm{Y,rec}} = h\nu_{\mathrm{Y}}\iiint_{\rm{T<T_{\mathrm{thr}}}} n_{\mathrm{e}}n_{\mathrm{H_{II}}}\alpha_{\mathrm{eff}}\left(T\right)\, \mathrm{d} V,
\end{equation}
and
\begin{equation} \label{eq:L_col}
\mathcal{L}_{\mathrm{Y,col}} = h\nu_{\mathrm{Y}}\iiint_{\rm{T<T_{\mathrm{thr}}}} n_{\mathrm{e}}n_{\mathrm{H_{I}}}\beta_{\mathrm{eff}}\left(T\right)\, \mathrm{d} V.
\end{equation}
Here, $h$ is Planck's constant, $n_{\rm e}$ the electron number density, $Y$ denotes Ly$\alpha$ or H$\alpha$, and $\nu_{\rm Y}$ the corresponding line frequency.
The total luminosity was then defined as
\begin{equation} \label{eq:L_totl}
\mathcal{L}_{\mathrm{Y}} = \mathcal{L}_{\mathrm{Y,col}} + \mathcal{L}_{\mathrm{Y,rec}} .
\end{equation}
\begin{figure}
    \centering
    \includegraphics[width=.99\linewidth]{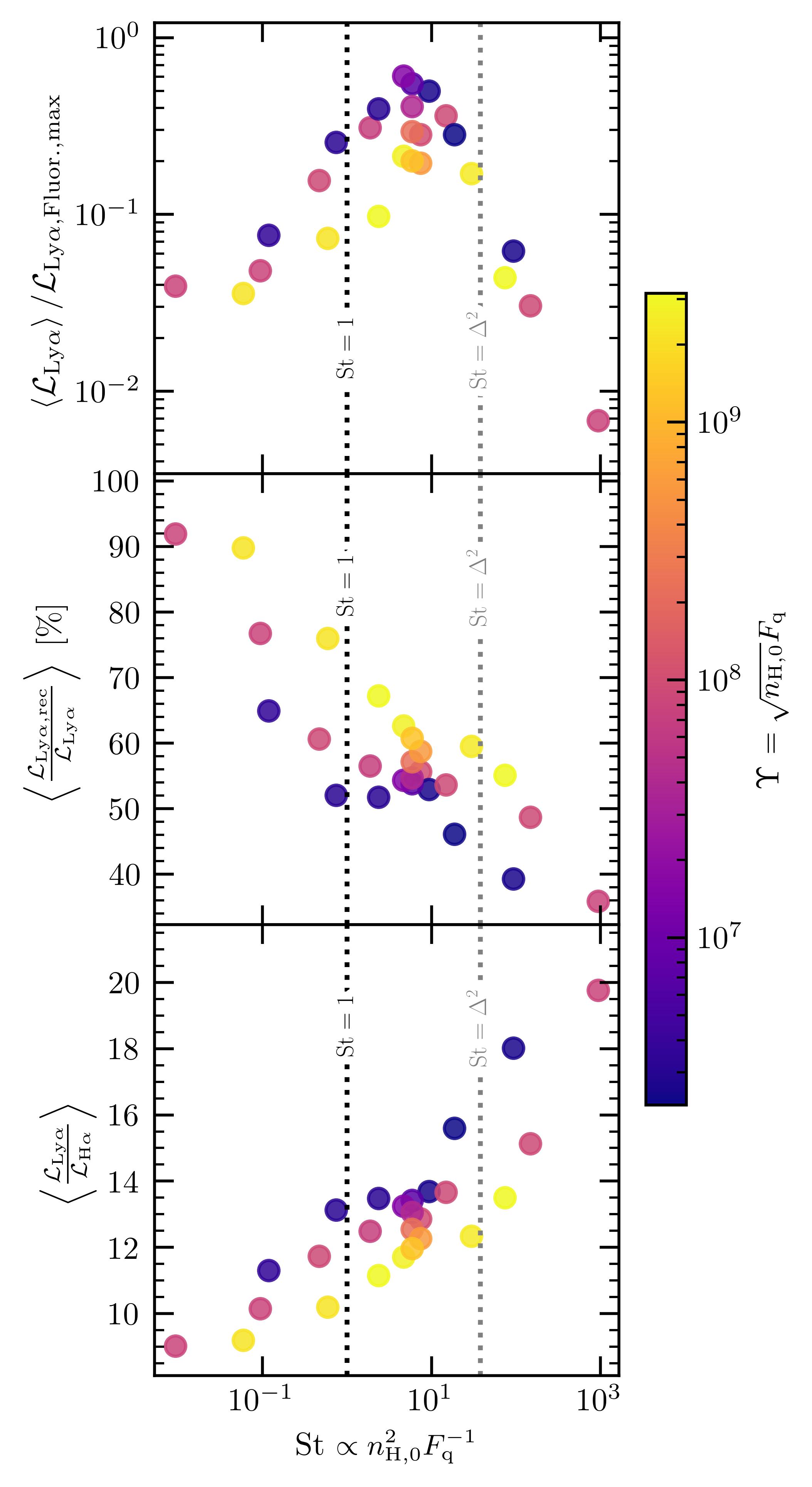}\caption{Emission properties of the clouds as a function of $\mathrm{St}$ (x-axis) and $\Upsilon$ (colour map). The circles represent time-averaged simulation values. {\it Top panel:} Ly$\alpha$ luminosity. {\it Middle panel:} Ratio of Ly$\alpha$ luminosity produced by recombination processes, expressed as a percentage. {\it Bottom panel :} Ly$\alpha$ to H$\alpha$ ratio.}
    \label{fig:H_lum}
\end{figure}
\\
The resulting luminosities averaged over $t\in[5,20]\,\rm Myr$ are shown in Fig.~\ref{fig:H_lum}. The Ly$\alpha$ luminosity was normalised by the maximum fluorescent luminosity,
\begin{equation} \label{eq:L_F_max}
\mathcal{L}_{\mathrm{Ly\alpha,Fluor.,max}} = h\nu_{\mathrm{Ly\alpha}}n_{\mathrm{H,0}}^2\alpha_{\mathrm{eff,Ly\alpha}}\frac{4\pi r_{\mathrm{c,0}}^3}{3},
\end{equation}
corresponding to a fully ionised initial cloud, which removes the trivial density and volume scaling.
For the optically thin regime, the analytical expectation for pure recombination is that the normalised luminosity decreases by $\sim \Delta^{-2}$ due to the quadratic dependence on density, and increases by $\sim \Delta$ owing to the larger emitting volume, yielding an overall scaling $\sim \Delta^{-1}$. Using $T_{\mathrm{i}} \sim 7\times 10^4\, \rm K$ from Fig.~\ref{fig:t_evo_cloud}, this gives a predicted value of $\sim 0.046$, in excellent agreement with Fig.~\ref{fig:H_lum}.

In the rocket-effect regime, the normalised luminosity increases markedly, reaching values up to two orders of magnitude higher than in other regimes. This enhancement is strongest for lower $\Upsilon$ and intermediate $\mathrm{St} \sim 6$. Here, $\mathrm{St} \sim 6$ marks the logarithmic midpoint between $\mathrm{St} = 1$ and $\mathrm{St} = \Delta^2$. Physically, this behaviour arises from the greater abundance of partially ionised hydrogen at lower $\Upsilon$ (see Fig.~\ref{fig:nH_pdf}, middle panel). In this regime, radiation does not penetrate efficiently through the PIR to ionise the remaining cloud. Consequently, the \ion{H}{II} gas within the PIR and the \ion{H}{I} gas near the ionisation front attain higher mean densities, enhancing both collisional and recombination emission.
\\
We then examined the fraction of luminosity arising from recombination (middle panel of Fig.~\ref{fig:H_lum}). This fraction decreases monotonically with increasing $\mathrm{St}$ and decreasing $\Upsilon$, as described by the following fit:
\begin{equation} \label{eq:r_rec_fit}
\left\langle\frac{\mathcal{L}_{\mathrm{Ly\alpha,rec}}}{\mathcal{L}_{\mathrm{Ly\alpha}}}\right\rangle  \sim 31.44 \times\mathrm{St}^{ -0.0704 } \Upsilon^{ 0.0383 },
\end{equation}
which yields a mean relative error of $0.5\,\%$. This trend holds across all regimes, with $\left\langle \mathcal{L}_{\mathrm{Ly\alpha,rec}} / \mathcal{L}_{\mathrm{Ly\alpha}} \right\rangle$ varying between $\sim 36-92\, \%$. In the rocket-effect regime, a more rapid ionisation of the PIR, occurring for higher $\Upsilon$ and lower $\mathrm{St}$, enhances the recombination contribution. A similar trend is observed in the radiation-shielded regime, where a higher ionisation strength parameter $\Upsilon$ likewise boosts the recombination contribution.
We stress, however, that the contribution of collisional excitation to Ly$\alpha$ emission in our results should be considered as an upper limit in case of non-primordial metallicities. We provide further discussion on this in Sect.~\ref{sec:caveats}. 
\\
Next, we considered the $\rm Ly\alpha / H\alpha$ line ratio (bottom panel). As with the recombination contribution, the line ratio follows a monotonic trend, increasing with higher $\mathrm{St}$ and lower $\Upsilon$, and is described by the following fit,
\begin{equation} \label{eq:r_Lya_Ha_fit}
\left\langle\frac{\mathcal{L}_{\mathrm{Ly\alpha}}}{\mathcal{L}_{\mathrm{H\alpha}}}\right\rangle  \sim 19.97 \times\mathrm{St}^{ 0.0592 } \Upsilon^{-0.0293}.
\end{equation}
which yields a mean relative error of $0.28\,\%$.
The relative contributions of recombination and collisional processes primarily drive the variation in the line ratio. First, the ratio for the effective recombination rate (considering case~B) varies only weakly $h\nu_\mathrm{Ly\alpha}\alpha_{\mathrm{eff,Ly\alpha}} / h\nu_\mathrm{H\alpha} \alpha_{\mathrm{eff,H\alpha}}\sim 6.6-8.3$ for $T\in[5000,10^{5.5}]\, \rm K$. By contrast, over the same temperature range, the ratio of effective collisional rates spans a much broader range, $h\nu_\mathrm{Ly\alpha}\beta_{\mathrm{eff,Ly\alpha}} / h\nu_\mathrm{H\alpha}\beta_{\mathrm{eff,H\alpha}}\sim 902-20.5$ (see Appendix~\ref{app:eff_rates}). In the optically thin regime, the line ratio is roughly $\in [8,14]$ due to the dominance of recombination. In the other regimes, the contribution of collisional excitation becomes more significant. The increase in the line ratio is then due to the lower hydrogen gas temperatures attained at lower $\Upsilon$ and higher $\mathrm{St}$ (see Fig.~\ref{fig:t_evo_cloud}), which favour higher values of $\beta_{\mathrm{eff,Ly\alpha}} / \beta_{\mathrm{eff,H\alpha}}$. We emphasise that the $\mathrm{Ly}\alpha/\mathrm{H}\alpha$ ratios reported here are intrinsic to the cloud. As Ly$\alpha$ is a resonant line, unlike H$\alpha$, photons are liable to scatter before escape, thereby reducing the observed Ly$\alpha$ surface brightness and the inferred line ratio \citep[e.g.][]{Leibler2018,Langen2023}, unless the observational aperture is sufficiently large to encompass the scattered emission.
\\
We further provide alternative fits in terms of the ionisation parameter $\mathcal{U}=F_{\mathrm{q}}/(n_{\mathrm{H}}c)$ and the PIR Strömgren number $\mathrm{St}/\Delta^2$ in Appendix~\ref{app:fits4obs}.

\section{Discussion}\label{sec:discussion}
The results presented in Sect.~\ref{sec:results} show that quasar radiation can significantly alter the physical properties of cold CGM clouds. Beyond modifying their ionisation and temperature through photoionisation and photo-heating, radiation also reshapes their morphology and density structure. This latter effect is particularly relevant for interpreting Ly$\alpha$ emission from quasar-illuminated haloes. Moreover, we find that the evolution of illuminated clouds can be captured by a small set of analytical parameters, primarily $\mathrm{St}$ and $\Upsilon$.
\\
Having validated our analytical framework for individual clouds, we extend it to a statistical ensemble in order to describe the propagation of quasar radiation through the CGM. We complement this approach with an upper-limit scenario corresponding to a continuous cold stream. Finally, we discuss how additional physical processes may modify our conclusions. As the main parameter for predicting the cloud regime is $\mathrm{St}$, we here limit our discussion to this parameter.

\subsection{Analytical model for the halo}\label{sec:model_halo}
\subsubsection{Model for a population of clouds}
We introduced a simple analytical model to derive a global parameter, $\mathrm{St}_{\mathrm{l}}$, that characterises the cumulative impact of multiple clouds on radiation along a ray originating from the quasar. The full derivation is provided in Appendix~\ref{app:model_halo}.
We assumed that the density of cold gas follows a power-law profile,
\begin{equation}\label{eq:nH_c}
    n_{\mathrm{H,c}}\left(r\right) \sim n_0 \left(\frac{r}{r_0}\right)^{-a_{\mathrm{n}}},
\end{equation}
where $r_0=10\, \rm pkpc$ defines a fiducial inner radius for the CGM, $n_0$ is the hydrogen number density at $r_0$, and $a_{\mathrm{n}}=5/4$ is inferred from observations of giant Ly$\alpha$ nebulae around quasars with MUSE \citep{Pezzulli2019}.

We assumed that the cold clouds occupy a fraction of the volume (or filling factor) $f_\mathrm{V}$, for which we adopted a fiducial value $f_\mathrm{V}=0.01$. Furthermore, we assumed that the cloud size is proportional to the local Jeans length,
\begin{equation}\label{eq:Rc}
    r_{\mathrm{c}}\left(r\right) = 0.5b \lambda_{\mathrm{J}}(r),
\end{equation}
where $b<1$ is a dimensionless proportionality factor, and $\lambda_{\mathrm{J}}(r)$ is the Jeans length.
For $b=0.1$ and $n_0=1\,\rm cm^{-3}$, this corresponds to cloud radii of approximately $80$ and $280\,\mathrm{pc}$ at halo-centric radii of $r=10$ and $75\,\mathrm{pkpc}$, respectively.

We considered the distribution of clouds along a line of sight originating from the quasar. The corresponding line-distribution, i.e. the number of clouds per unit radial path length, is
\begin{equation}\label{eq:d_l}
d_{\mathrm{l}}\left(r\right)=\frac{3f_{\mathrm{V}}}{4r_{\mathrm{c}}\left(r\right)},
\end{equation}
The mean number of clouds intersecting the line of sight out to radius $r$ is then,
\begin{equation}\label{eq:Nc_l}
    N_{\mathrm{c,l}}\left(r\right) =\int^r_{r_0} d_{\mathrm{l}}\left(r'\right) \,\mathrm{d}r' \propto \left(\frac{r}{r_0}\right)^{1-a_{\mathrm{n}}/2}-1.
\end{equation}
We defined the cumulative parameter for the Strömgren number,
\begin{equation}\label{eq:St__l}
     \mathrm{St}_{\mathrm{l}}\left(r\right) = \int^r_{r_0} \mathrm{St}\left(r'\right) d_{\mathrm{l}}\left(r'\right) \,\mathrm{d}r' \propto \left(\frac{r}{r_0}\right)^{3-2a_{\mathrm{n}}}-1,
\end{equation}
and its mean value,
\begin{equation}\label{eq:St__ml}
     \left\langle \mathrm{St} \right\rangle_{\mathrm{l}}\left(r\right) =\frac{\mathrm{St}_{\mathrm{l}}\left(r\right)}{N_{\mathrm{c,l}}\left(r\right)}.
\end{equation}
The integral Strömgren number $\mathrm{St}_{\mathrm{l}}$ measures the cumulative interaction between radiation and cold gas along the line of sight. Notably, the $\mathrm{St}_{\mathrm{l}}$ is independent of the individual cloud size and of our choice of $r_{\mathrm{c}}\propto\lambda_{\mathrm{J}}(r)$, meaning that the contribution of numerous small clouds is the same as that of a few large clouds.
The parameter $\left\langle \mathrm{St} \right\rangle_{\mathrm{l}}$ represents the mean value of $\mathrm{St}$ for all clouds encountered along a line of sight out to radius $r$.

Physically, $\mathrm{St}_{\mathrm{l}}$ plays a role analogous to $\mathrm{St}$ for a single cloud:
\begin{itemize}
    \item If $\mathrm{St}_{\mathrm{l}}(r)<1$, radiation freely propagates and ionises the CGM out to radius $r$.
    \item If $\mathrm{St}_{\mathrm{l}}(r)>\Delta^2$, the cold CGM becomes effectively radiation-shielded, and the radiation cannot propagate beyond a radius $r$.
    \item In the halo-rocket-effect regime, $1<\mathrm{St}_{\mathrm{l}}(r)<\Delta^2$, an ionisation front accompanied by a compressive wave propagates through successive clouds.
\end{itemize}

\subsubsection{Halo model results}
\begin{figure}
    \centering
    \includegraphics[width=.9\linewidth]{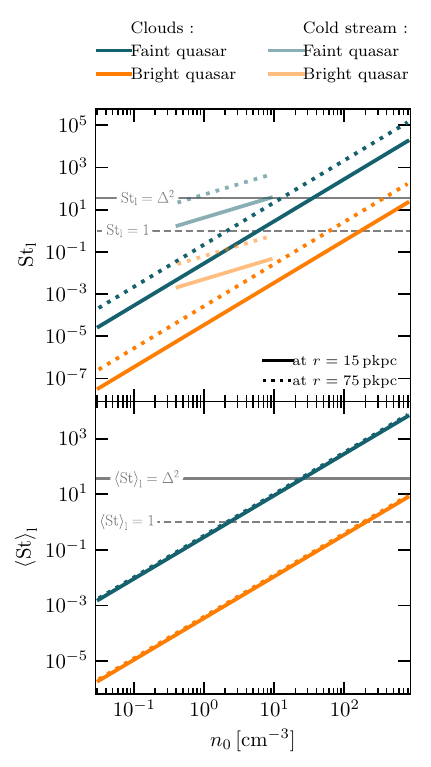}    
    \caption{Properties of a cloud ensemble as a function of the hydrogen number density at $r_0=10\,  \mathrm{pkpc}$ (i.e. $n_0$), with an assumed $f_\mathrm{V}=0.01$. Results are shown for faint (teal) and bright (orange) quasars at halo-centric radii of $r=15$ (solid lines) and $75\, \rm kpc$ (dotted lines). The quasar ionising photon rates correspond to Lyman-limit specific luminosities of $L_{\mathrm{\nu,LL}} \sim 10^{28.6}$ and $10^{31.6} \, \rm erg\, s^{-1}\, Hz^{-1}$, representative of faint quasars \citep{Mackenzie2021} and bright quasars \citep{Borisova16}. {\it Top panel:} Cumulative parameter $\mathrm{St}_{\mathrm{l}}$, determining if the ionisation front, up to a radius $r$, smoothly ionises the cold gas ($\mathrm{St}_{\mathrm{l}}<1$), compresses the neutral gas ahead leading to a rocket-effect ($\mathrm{St}_{\mathrm{l}}\in[1,\Delta^2]$), or remains stalled ($\mathrm{St}_{\mathrm{l}}>\Delta^2$). The parameter is shown for both a ray crossing a cloud population and a ray propagating inside a cold stream. The latter case is described in Appendix~\ref{app:St_s}.
    {\it Bottom panel:} Mean $\left\langle \mathrm{St}\right\rangle_{\mathrm{l}}$ parameter for all clouds within a radius $r$, using a Jeans proportionality factor $b=0.1$ (Eq.~\ref{eq:Rc})}
    \label{fig:St_l}
\end{figure}
Figure~\ref{fig:St_l} presents the cumulative parameter $\mathrm{St}_{\mathrm{l}}$ and the mean cloud parameter $\langle\mathrm{St}\rangle_{\mathrm{l}}$ for quasar luminosities representative of faint \citep[e.g.][]{Mackenzie2021} and bright quasars \citep[e.g.][]{Borisova16}. Results are shown for two halo-centric radii: $r=15\,\mathrm{pkpc}$, representative of the inner CGM, and $r=75\,\mathrm{pkpc}$, which roughly corresponds to their virial radius \citep{Pezzulli2019}. In addition to the cloud population, we include an indicative upper limit for $\mathrm{St}_{\mathrm{l}}$ computed for one ray propagating entirely within a cold stream in a $10^{12}\,\rm M_\odot$ halo at $z=3$. This configuration effectively corresponds to a continuous medium along the line of sight ($f_{\mathrm{V}}=1$), and therefore maximises the interaction between radiation and cold gas. Details of this scenario are provided in Appendix~\ref{app:St_s}. The cold-stream results are shown only within the range of $n_0$ permitted by the empirical toy-model of \citet{Mandelker2020b}.
For our assumed $f_\mathrm{V}=0.01$, the maximum values of $n_0$ for which radiation can still escape the CGM ($\mathrm{St}_{\mathrm{l}}<\Delta^2$) are approximately $12$ and $373\,\mathrm{cm^{-3}}$ for faint and bright quasars, respectively. The maximum $n_0$ for which the radiation field can fully ionise the cold gas ($\mathrm{St}_{\mathrm{l}}<1$) are approximately $2.13$ and $61\,\mathrm{cm^{-3}}$ for faint and bright quasars. For cold streams ($f_{\mathrm{V}}=1$), bright quasars fully ionise the gas, while faint quasars produce a propagating ionisation front that stalls within the CGM. However, given the small covering fraction of such streams, most ionising radiation is expected to escape into the IGM.
\\
The mean parameter $\langle\mathrm{St}\rangle_{\mathrm{l}}$ varies only weakly with radius. For both faint and bright quasars with $n_0 \lesssim 1\,\rm cm^{-3}$, the mean number of clouds intersected along a line of sight out to $r=75\,\rm pkpc$ is below unity. In this regime, cumulative effects remain limited and $\mathrm{St}_{\mathrm{l}} \lesssim \langle \mathrm{St} \rangle_{\mathrm{l}}$, reflecting a sparse population of relatively large clouds. The resulting emission is therefore expected to be anisotropic and dominated by recombination processes (see Fig.~\ref{fig:H_lum}).
The rocket-effect regime and its associated enhancement of Ly$\alpha$ emission are primarily relevant for faint quasars, while brighter quasars typically drive the system towards the optically thin regime. Although unlikely, a larger $f_{\mathrm{V}}$ or more massive haloes hosting dense cold streams could shift bright quasar environments into the rocket-effect regime (see Fig.~\ref{app:St_s}). However, as explained below, this would also imply luminosities larger than observed.
Furthermore, considering a cold structure of density $n_{\mathrm{H}}=10\,\rm cm^{-3}$ and characteristic size $L_0\sim 0.1\,\rm kpc$ in the IGM around a bright quasar, the distance at which it enters the rocket-effect regime is $d\sim 2\,\mathrm{Mpc} \propto (n_{\mathrm{H}}\sqrt{L_0})^{-1}$ (Eq.~\ref{eq:St}). This suggests that bright quasars are likely to fully ionise most of their surrounding IGM.

Importantly, $n_0$ denotes the pre-illumination density, whereas observational constraints probe post-ionisation conditions. To enable comparison, pre-ionisation densities must be converted using a factor $\Delta^{-1}$. 
For bright quasars, we find that the minimum post-ionisation density required at $10\,\rm pkpc$ to enter the rocket-effect regime is $\sim 10\,\rm cm^{-3}$. For comparison, \citet{Pezzulli2019} infer a post-ionisation density of $n_0 = 3.8\,\rm cm^{-3}$ at the same radius to reproduce the Ly$\alpha$ surface brightness of nebulae around bright quasars \citep{Borisova16}, assuming $f_\mathrm{V}=0.01$ (as in our model) and $T = 7 \times 10^4\,\rm K$ (Fig.~\ref{fig:t_evo_cloud})\footnote{We emphasise that both the observed surface brightness and the integrated $\mathrm{St}$ depend on $f_\mathrm{V} n_0^2$, so the conclusion that observed nebulae around bright QSOs are in the thin regime is independent on the assumed value of $f_{\mathrm{V}}$.}.
This indicates that bright quasar nebulae are consistent with the fully ionised, optically thin regime (see also Sect.~5.1 of \citet{Pezzulli2019}). On the other hand, we note that for faint quasars, more prone to the rocket-effect regime, a formalism such as that in \citet{Pezzulli2019} could be applied only after correcting for a fraction of recombination-to-total luminosity (typically $50-60 \%$, Fig.~\ref{fig:H_lum}) and for unresolved internal structure. This is quantified in our case by the clumping factor $\mathcal{C}_{\mathrm{int}}\sim \langle \mathcal{C}_{\mathrm{H_{II},sim}}\rangle \lesssim 2$ (Figure \ref{fig:clumping_f}), although we cannot exclude that additional clumpiness might be induced by other (also non-radiation-hydrodynamics) effects currently neglected in this work (see Sect. \ref{sec:caveats}).

The present model suits a simplified description of a relatively isotropic cold CGM with constant $f_{\mathrm{V}}$ and a single characteristic cloud size at each radius. In reality, the CGM likely exhibits a broad distribution of densities and sizes, which may alter the detailed propagation of ionisation fronts. In Appendix~\ref{app:St_s}, we further explore the case of highly anisotropic structures such as cold streams. Future extensions incorporating density distributions and radially varying filling factors will enable more realistic predictions for Ly$\alpha$ emission in quasar haloes.

\subsection{Additional physics}\label{sec:caveats}
We now consider the potential impact of additional physical processes.
\paragraph{Dynamics and halo potential:} Our simulations assumed static clouds to isolate radiative effects. In the CGM, however, cold structures are expected to move within the halo potential. We provide our estimation of the acceleration of the cometary globule in the rocket-effect regime in Appendix~\ref{app:acc_CG}. For a typical $10^{12}\,\rm M_\odot$ halo at $z=3$, this simple estimate indicates that the rocket-effect acceleration initially exceeds the gravitational acceleration at radii $r \gtrsim 33\,\rm pkpc$. However, as the globule does not retain its initial structure and may undergo ionisation or fragmentation, this acceleration likely persists only for a few megayears. Together with hydrodynamical instabilities arising from the relative motion with the hot background \citep[e.g.][]{Armillotta2017,Gronke2018}, these effects may further modify the cloud evolution and warrant more detailed investigation in future work.
\paragraph{Metal cooling:} As the ionisation front is governed primarily by hydrogen, the inclusion of metals does not qualitatively alter the framework of Sect.~\ref{sec:theory}. Metal-line cooling would lower the temperature of the PIR, leading to higher densities in the cloud and cometary globule. However, \citet{Nakatani2019} show that even for denser ISM clouds, metallicities $Z \lesssim 10^{-2}\,\rm Z_\odot$ do not significantly affect the dynamics. While metals are therefore unlikely to modify our main conclusions, they could reduce the contribution of collisional excitation to Ly$\alpha$ emission by providing alternative cooling channels. A quantitative assessment would require a full chemical network, which is beyond the scope of this work.
\paragraph{Magnetic field:} Magnetic fields have been shown \citep{bertoldi_photoevaporation_1989,bertoldi_photoevaporation_1990} to inhibit the compression of cold gas, thereby reducing the density attained by the cometary globule compared to purely hydrodynamical cases. This effect applies primarily when the initial plasma parameter $\beta = P/P_{\mathrm{mag}} \sim 1$, with $P_{\mathrm{mag}}$ the magnetic pressure. Such conditions may occur for relatively small CGM clouds \citep{Nelson2020} or within the mixing layers of cold inflows \citep{Ledos2024a,Kaul2025}.
\paragraph{Radiation pressure:} The radiation pressure may be estimated as $P_{\mathrm{rad}}\sim L_{\mathrm{q,EUV}} /(4\pi r^2 c)$. Compared to the initial thermal pressure, we find $P_{\mathrm{rad}}/P \lesssim 2.6$, with peak values in the optically thin regime. In practice, however, once the gas becomes fully ionised, most photons escape and the effective radiation pressure drops, yielding $P_{\mathrm{rad}}/P < 1$. Radiation pressure, therefore, plays a secondary role under the conditions explored here, and may become significant only under substantially higher photon fluxes.
\paragraph{Self-gravity:} In star-forming regions, compression by ionisation fronts can trigger gravitational collapse \citep[e.g.][]{Sugitani1991,Sugitani1994,Sugitani1995,Megeath1996,lefloch_cometary_1994,Nakatani2019}. As our simulations neglect self-gravity, metal-line, and molecular cooling, they likely overestimate temperatures and underestimate densities. Assuming a cometary globule of radius $\sim r_{\mathrm{c}}/4$, density $n_{\mathrm{H,CG}}\sim 5 n_{\mathrm{H,0}}$, and temperature $T_{\mathrm{CG}}\sim T_{\mathrm{c,0}}/2$, we find that gravitational instability occurs when $2r_{\mathrm{c}}\gtrsim 2\lambda_{\mathrm{J,0}}/3$, corresponding to $b\gtrsim0.33$ in Eq.~\ref{eq:Rc}. Thus, sufficiently large structures in the rocket-effect regime may undergo collapse and potentially form stars. A detailed treatment is left to future work.

Additional mechanisms such as Compton heating by X-rays, thermal conduction, and cosmic-ray interactions may also influence cloud evolution. They warrant dedicated future studies.

\section{Summary}\label{sec:ccl}
Quasar radiation at high redshift has been successfully used over the past decades to enhance the detectability of CGM emission and to provide new constraints on CGM physical properties. In this work, we presented an analytical framework, supported by simulations, to predict under which conditions quasar radiation significantly affects CGM structures and their emission.  

We first introduced an analytical framework (Sect.~\ref{sec:theory}) describing cloud evolution in terms of two independent parameters: $\mathrm{St}$, the Strömgren number describing the balance between ionising photons and recombination, and $\Upsilon$, an ionisation strength parameter. In addition, we defined an effective threshold, $\mathrm{St}=\Delta^2$ with $\Delta$ the ratio of the post-ionisation to pre-ionisation cold-gas hydrogen number density, for the onset of radiation-shielding, taking into account the hydrodynamical response of the gas to photoionisation. This framework was validated against a suite of radiation-hydrodynamic simulations of static cold clouds illuminated by EUV photons, from which we further analysed the resulting emission properties (Sect.~\ref{sec:results}). Finally, we extended the single-cloud analytical framework to an ensemble of clouds and a single cold stream (Sect.~\ref{sec:model_halo}), providing a self-consistent description of radiation penetration through a population of cold structures within a hot CGM. Our main conclusions are as follows:
\begin{itemize}
    \item Ionisation front propagation and density structure: Even in the absence of radiation pressure, the propagation of an ionisation front through a cloud initially at uniform density produces an ionised cloud with a non-uniform density distribution. This effect is particularly strong in the rocket-effect regime, where quasar radiation heats the illuminated side, compresses the shadowed side, and accelerates the residual cold clump, forming a cometary globule (Fig.~\ref{fig:model_illust}).
    \item Ly$\alpha$ emission: Clouds in the rocket-effect regime exhibit the highest Ly$\alpha$ luminosities, up to one order of magnitude higher than fully ionised clouds (Fig.~\ref{fig:H_lum}). This enhancement is driven by the gradual ionisation of dense gas at the advancing ionisation front, with roughly $60\%$ of the luminosity arising from recombination. Both the recombination fraction and the Ly$\alpha$/H$\alpha$ line ratio are well described by power-law fits as functions of $\mathrm{St}$ and $\Upsilon$, or alternatively as functions of the ionisation parameter $\mathcal{U}$ and $\mathrm{St}/\Delta^2$.
    \item Halo scale effects: Using the cumulative parameter $\mathrm{St}_{\mathrm{l}}$, we find that the cold CGM around bright quasars ($L_{\mathrm{\nu,LL}} \sim 10^{31.6} \, \rm erg\, s^{-1}\, Hz^{-1}$) is likely fully ionised, whereas the CGM around faint quasars ($L_{\mathrm{\nu,LL}} \sim 10^{28.6} \, \rm erg\, s^{-1}\, Hz^{-1}$) predominantly experiences a rocket-effect regime. This is particularly relevant for dense clouds and cold streams around faint quasars. 
    A higher cold gas fraction in our model, dense cold structures in the IGM, or larger and more massive haloes could induce a rocket-effect regime even around bright quasars, although this would imply luminosities larger than observed. 
\end{itemize}
Overall, our work provides a predictive, analytically grounded framework, validated by simulations, for assessing the impact of quasar ionising radiation on cold CGM structures. These processes are likely unresolved or absent in current cosmological simulations due to limited resolution and the lack of on-the-fly radiative transfer. Future work will translate this framework into a subgrid prescription for cosmological simulations, enabling robust predictions for Ly$\alpha$ surface-brightness distributions.

\begin{acknowledgement}
We acknowledge Saeed Sarpas for insightful discussions at the early stages of this project, which was partly inspired by his PhD thesis \citep{Saeed2021}. N.L. and S.C. acknowledge the support of the European Research Council (ERC) under the European Union’s Horizon 2020 research and innovation program grant agreement No 864361. We acknowledge the CINECA award under the ISCRA initiative for the availability of high-performance computing resources and support. A.P. acknowledges support from the Independent Research Fund Denmark (DFF) under grant 3120-00043B.
\end{acknowledgement}

\bibliographystyle{bibtex/aa} 
\bibliography{references.bib} 

\begin{appendix}
\section{Ionisation front}\label{app:theory}
In parallel with Section~\ref{sec:theory}, which focuses on the derivation of the cloud's regimes, we hereby provide a more detailed description of the ionisation fronts, as well as a second derivation of our analytical framework based on the ionisation front point of view.

\subsection{Description and classification}\label{sec:sub:i_front}
Here, we describe the general dynamics of the ionisation front propagation. When exposed to radiation, an ionisation front will form in the cold gas. Following \citet{spitzer_physical_1978}, particle conservation across the front implies,
\begin{equation}\label{eq:ni_nc_Fq}
    n_{\mathrm{H.i}}u_{\mathrm{I,i}} \sim n_{\mathrm{H,c}}u_{\mathrm{I,c}} \sim F_{\mathrm{q}},
\end{equation}
where $n_{\mathrm{H,*}}$ and $u_{\mathrm{*}}$ denote the hydrogen number density and velocity on each side of the ionisation front, the subscript i and c standing for the ionised and the cold phases.
Along with energy conservation and the perfect gas law, one obtains the jump condition
\begin{equation}\label{eq:jump_dens}
    \frac{\rho_{\mathrm{i}}}{\rho_{\mathrm{c}}}= \left\{\begin{array}{l}
    \frac{\left(u_{\mathrm{I}}^2 + c_{\mathrm{s,c}}^2\right) \pm \left[\left(u_{\mathrm{I}}^2 + c_{\mathrm{s,c}}^2\right)^2 - 4u_{\mathrm{I}}^2c_{\mathrm{s,i}}^2 \right]^{1/2} }{2c_{\mathrm{s,i}}^2} \, \, (\text{dual roots})\\
    \frac{u_{\mathrm{I}}^2 + c_{\mathrm{s,c}}^2}{2c_{\mathrm{s,i}}^2}  \, \, (\text{unique root}),
    \end{array} \right.  
\end{equation}
where $u_{\mathrm{I}}$ is the velocity of the ionisation front and $c_{\mathrm{s,*}}$ is the isothermal sound speed in each phase.
The condition for the dual roots solution leads to the characteristic velocities,
{\small
\begin{equation}
    \left\{ \def\arraystretch{0.5}\begin{array}{ll}
    u_{\rm{R}} = &  c_{\mathrm{s,i}} + \left[ \left(c_{\mathrm{s,i}}^2-c_{\mathrm{s,c}}^2\right) \right]^{1/2}\\
    u_{\rm{M}} = &  c_{\mathrm{s,i}} \\
    u_{\rm{D}} = & c_{\mathrm{s,i}} - \left[ \left(c_{\mathrm{s,i}}^2-c_{\mathrm{s,c}}^2\right) \right]^{1/2}\\
    \end{array}\right.
\end{equation}}
which define the different regimes of ionisation-front propagation. If $u_{\rm{I}} > u_{\rm{R}}$, the front is of R-type. Such fronts occur for sufficiently high photon fluxes and propagate supersonically with respect to the gas. R-type fronts can appear in strong or weak forms corresponding to the $+$ and $-$ branches of equation~\ref{eq:jump_dens}, respectively. If $u_{\rm{D}} < u_{\rm{I}} < u_{\rm{R}}$, the front becomes M-type, which also leads to the propagation of a shock in the cold phase with slight variations depending on whether $u_{\mathrm{I}}$ is larger or smaller than $u_{\mathrm{M}}$. Finally, if $u_{\rm{I}} < u_{\rm{D}}$, the front is of D-type, "D" referring to dense gas as in this case $\rho_{\mathrm{c}}<\rho_{\mathrm{i}}$.

If $u_{\rm{i}} = u_{\rm{D}}$ or $u_{\rm{i}} = u_{\rm{R}}$ or $u_{\rm{i}} = u_{\rm{M}}$, we obtain then critical states which are described by the unique root solution of equation~\ref{eq:jump_dens}.

\subsection{Alternative derivation of the model}\label{app:sub:St2}
Taking the same initial set-up as in Section~\ref{sec:theory}, we can first define for the initial neutral cloud the recombination time,
\begin{equation}\label{eq:t_rec}
    t_{\mathrm{rec}} = \frac{1}{n_{\mathrm{H,0}} \alpha^{\mathrm{B}}_{\mathrm{H}}}.
\end{equation}
The ionisation front velocity within the cloud \citep{spitzer_physical_1978} is,
\begin{equation}
    u_{\mathrm{I,c}} = F_{\mathrm{q}}/n_{\mathrm{H,0}},
\end{equation}
which yields an ionisation-front crossing time in the cloud of
\begin{equation}
    t_{\mathrm{I,c}} = \frac{2r_{\mathrm{c,0}}}{u_{\mathrm{I,c}}}.
\end{equation}
By taking the ratio of the two timescales, we then recover the Str\"{o}mgren number, 
\begin{equation}\label{eq:St_2}
    \mathrm{St} \equiv \frac{t_{\mathrm{I,c}}}{t_{\mathrm{rec}}} = \frac{2r_{\mathrm{c,0}}\alpha^{\mathrm{B}}_{\mathrm{H}}n_{\mathrm{H,0}}^2}{F_{\mathrm{q}}}.
\end{equation}
To recover the new radiation-shielded regimes, one can then express the two latter timescales for the PIR, leading to,
\begin{equation}\label{eq:St_i_2}
    \frac{t_{\mathrm{I,i}}}{t_{\mathrm{rec,i}}} = \frac{2r_{\mathrm{c,0}}\alpha^{\mathrm{B}}_{\mathrm{H}}n_{\mathrm{H,i}}^2}{F_{\mathrm{q}}} = \mathrm{St}\times\left(\frac{n_{\mathrm{H,i}}}{n_{\mathrm{H,0}}}\right)^2= \mathrm{St}\times\Delta^{-2}.
\end{equation}

\section{Numerical verification}\label{app:GC}
\subsection{Grid convergence study}
The fidelity of the simulation suite is assessed through a verification procedure inspired by classical grid-convergence studies \citep{Steffen1995,Roache1998}. Each simulation is initially performed with  AMR refinement levels of 6, 7, and 8.
We define a selected diagnostic quantity $X_l$ (e.g. the clumping factor), with $l$ denoting the level of refinement of the chosen simulation. The mesh is refined wherever sufficiently sharp spatial variations develop. Specifically, refinement is triggered when the normalised gradient between neighbouring cells exceeds 0.1 in the gas density, pressure, \ion{H}{I} fraction, or \ion{H}{II} fraction \citep[equation~46]{Rosdahl2015}.

For each simulation, we compute the relative error between the highest refinement level $l_{\rm max}$, initially 8, and the lower levels as,
\begin{equation} \label{eq:err_conv}
\epsilon_l\left(X\right) \equiv \left\vert \frac{ X_l - X_{l_{\mathrm{max}}} }{X_{l_{\mathrm{max}}}} \right\vert.
\end{equation}
Each additional refinement level corresponds to a minimum grid spacing $\Delta x$ getting divided by 2. Using the values of $X_l$ obtained at successive AMR levels, we compute the theoretical linear extrapolation of the fully converged solution for $\Delta x \rightarrow 0$, i.e. the Richardson extrapolation,
\begin{equation} \label{eq:rich_conv}
X_{l_\infty} \equiv X_{l_{\mathrm{max}}} + \frac{X_{l_{\mathrm{max}}}-X_{l_{\mathrm{max}}-1}}{2^p-1},
\end{equation}
where $p=2$ is the nominal spatial and time order of accuracy of RAMSES-RT.
The estimated discretisation error at refinement level $l_{\rm max}$ is then given by,
\begin{equation} \label{eq:err_conv_inf}
\epsilon_{l_\infty}\left(X\right) \equiv \left\vert \frac{X_{l_{\mathrm{max}}} - X_{l_\infty} }{X_{l_\infty}} \right\vert.
\end{equation}
We define two levels of convergence to assess simulation reliability:
\begin{itemize}
    \item {Weak criteria:} for a given simulation, the time-averaged relative errors of the cold-gas number density (shown in Fig.~\ref{fig:t_evo_cloud}) $\left\langle \epsilon_{l_{\mathrm{max}}}\left( n_{\mathrm{c}}\right) \right\rangle $ and $\left\langle \epsilon_{l_\infty}\left( n_{\mathrm{c}}\right) \right\rangle $ must both be below $10\,\%$, 
    \item {Strong criteria:} for a given simulation, the time-average relative error of the \ion{H}{II} clumping factor (shown in Fig.~\ref{fig:clumping_f}) $\left\langle \epsilon_{l_\infty}\left( \mathcal{C}_{\mathrm{H_{II},sim}}\right) \right\rangle$ must be below $10\,\%$.
\end{itemize}
The density-based criteria are termed weak because they rely on an integral quantity, $n_{\rm c}$, which typically exhibits good convergence even in the presence of shocks and in the absence of physically resolved dissipative processes.
By contrast, the clumping-factor criterion is designated strong, as this quantity is particularly sensitive to shocks and small-scale instabilities. In simulations without explicit physical viscosity, highly non-linear processes may amplify numerical instabilities \citep[see discussion of secondary instabilities in][]{McNally2012}. These instabilities can grow and hinder the convergence of local quantities \citep{Fidkowski2014} reflected by PDFs and clumping factors in our simulations. The strong criterion therefore, ensures the reliability of the simulations even with respect to fluctuations in local quantities.
\\
The weak criterion is applied to simulations in the optically thin and radiation-shielded regimes, whereas both weak and strong criteria are enforced in the rocket-effect regime. If a simulation does not satisfy its prescribed criteria, it is rerun with an additional refinement level. 
As a result, all simulations satisfy the weak criteria, with mean values $\left\langle \epsilon_{l_{\mathrm{max}}}\left( n_{\mathrm{c}}\right) \right\rangle \sim 4.5\,\%$ and $\left\langle \epsilon_{l_\infty}\left( n_{\mathrm{c}}\right) \right\rangle \sim 1.5\,\%$. Similarly, all simulations in the rocket-effect regime achieve the strong criterion with $\left\langle \epsilon_{l_\infty}\left( \mathcal{C}_{\mathrm{H_{II},sim}}\right) \right\rangle\sim 1\,\%$ and $\left\langle \epsilon_{l_{\mathrm{max}}}\left( \mathcal{C}_{\mathrm{H_{II},sim}}\right) \right\rangle\sim 2.86\,\%$. Among the 26 simulations, only 7 were run up to $l_{\mathrm{max}}=9$.
\\
We restrict the strong criterion to \ion{H}{II} for conciseness. Using \ion{H}{I}, for this criterion, we find two simulations, namely the one defined by $(\mathrm{St},\Upsilon)\sim (0.6,10^{9.3})$ and $(7.4,10^{8.8})$, with $\left\langle \epsilon_{l_{\infty}}\left( \mathcal{C}_{\mathrm{H_{I},sim}}\right) \right\rangle\sim 28$ and $52\,\%$, despite satisfying the weak criterion. The non-convergence of $\left\langle \mathcal{C}_{\mathrm{H_{I},sim}}\right\rangle$ for these two simulations does not seem to propagate to the other key quantities, thus likely not altering our conclusions.
Indeed, we also computed $\left\langle \epsilon_{l_{\mathrm{max}}} \right\rangle$ and $\left\langle \epsilon_{l_{\infty}} \right\rangle$ for the recombination and collisional luminosity of both Ly$\alpha$ and H$\alpha$ described in Section~\ref{sec:sub:emission}. The resulting mean value for $\left\langle \epsilon_{l_{\mathrm{max}}} \right\rangle$ and $\left\langle \epsilon_{l_{\infty}} \right\rangle$ among all luminosities are $8.7$ and $6.7\,\%$, respectively. Using any of the above luminosities for our strong criterion leads to all our simulations satisfying it.
\\
This verification procedure demonstrates a robust quantification of our simulations' convergence. In addition, the adaptive approach appears to be energy-efficient, as by refining only simulations that fail to meet the prescribed accuracy thresholds, we avoid systematically running all models at the highest AMR level, thereby reducing both computational cost and energy consumption. Thus, resolution is increased only where needed, and not enforced where it yields no further gain.

\subsection{Photon groups convergence study}\label{app:phot_conv}
The radiative transfer module of RAMSES-RT \citep{Rosdahl2015} discretises the radiation field into a finite number of photon groups over the adopted SED range of $1-100\,\rm Ryd$. However, the width $w_\mathrm{i-front}$ of an ionisation front depends on the photon mean free path \citep{Spitzer1962,Osterbrock2006},
\begin{equation}
    \lambda_\mathrm{mfp,ph} \sim \frac{1}{n_\mathrm{H_I}\sigma_\mathrm{H_I}(\bar{h\nu})}
\end{equation}
where $\sigma_\mathrm{H_I}(\bar{h\nu})$ is the \ion{H}{I} photoionisation cross-section evaluated at the characteristic mean photon energy $\bar{h\nu}\simeq p/(p-1)\nu_0\sim2.4\,\rm Ryd$.
For our simulations in the rocket-effect regime, the largest characteristic mean free paths occur in the lowest-density clouds ($n_\mathrm{H}\sim0.6-1\,\rm cm^{-3}$), yielding $\lambda_\mathrm{mfp,ph}\sim0.6-1\,\rm pc$. Since the width of an ionisation front may exceed the characteristic photon mean free path by up to one order of magnitude \citep{Shapiro2004}, the resulting front width can become comparable to the cloud size. In addition, higher-energy photons, owing to their larger mean free path, may penetrate beyond the main ionisation front and partially pre-heat otherwise neutral gas. 

To assess if our RAMSES-RT photon group discretisation is adequate, we first perform 1D simulations to study the variation of the temperature profile and the ionisation-front width with increasing number of photon groups $M$. We then investigate the impact of $M$ on the gas density PDFs in representative 3D simulations. 

\subsubsection{Ionisation front width check}
\begin{figure}
    \centering
    \includegraphics[width=1.\linewidth]{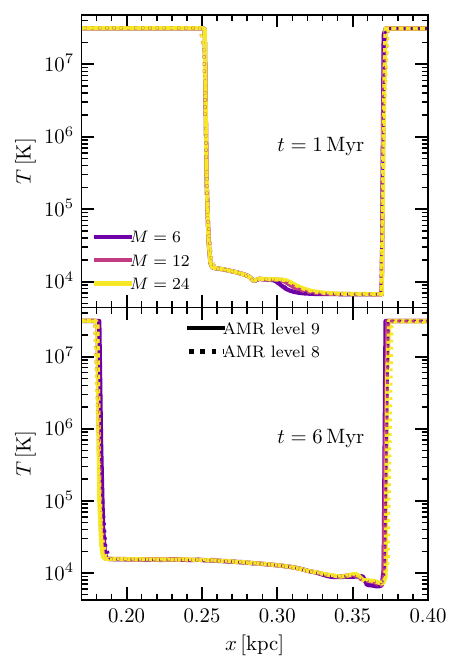}
    \caption{Temperature profile of the 1D test simulations at $t=1$ and $6\,\rm Myr$ for three different numbers of photon groups and two different resolutions.}
    \label{fig:Tprof_conv_M_photon_group}
\end{figure}

To investigate the impact of photon-group discretisation on the thermal and ionisation-front structures within the cloud, we performed additional 1D simulations with $M=6,12$ and $24$. These simulations isolate radiative broadening from multidimensional hydrodynamical effects.
\\
The initial conditions are identical to those of the first-column simulation shown in Fig.~\ref{fig:evo_cloud_ups}, with $n_\mathrm{H,0}\sim 1\,\rm cm^{-3}$ and $F_\mathrm{q}\sim 10^{6.4} \, \rm cm^{-2}\,s^{-1}$, corresponding to one of the lowest-density cases explored in the rocket-effect regime and therefore among those most susceptible to ionisation-front broadening and neutral gas pre-heating by hard photons.
The computational domain spans $x\in[0,0.64]\, \rm kpc$, with the cloud represented by a step-function density profile centred in the domain and an initial radius $r_\mathrm{c,0}=50\,\rm pc$.
The temperature profile within the cold gas is shown in Fig.~\ref{fig:Tprof_conv_M_photon_group} for $M=6,12$ and $24$, and 2 different resolutions ($l_\mathrm{max}=8$ and $9$).
The temperature profiles after $1\,\rm Myr$ show modest differences between spectral resolutions. In particular, the heated region extends approximately $10\,\rm pc$ deeper into the cloud in the $M=24$ run than in the $M=6$ run, although the overall profile shape remains unchanged. The $M=12$ and $24$ cases appear nearly converged. The discrepancies decrease with time, with temperature differences of only $\sim 10\%$ remaining by $t=6\,\rm Myr$. No differences are observed between the two spatial resolutions, indicating that the temperature structure is spatially resolved.

We then define the ionisation-front width $w_\mathrm{i-front}$ as the distance between the positions where the neutral fraction satisfies $f_\mathrm{H_I}= 0.1$ and $0.9$.
\begin{figure}
    \centering
    \includegraphics[width=1.\linewidth]{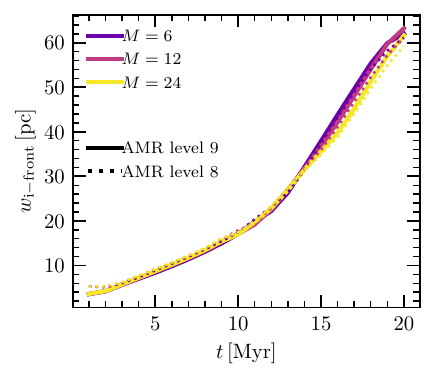}
    \caption{Time evolution of the ionisation-front width of the 1D test simulation for different numbers of photon groups $M$ and two different resolutions.}
    \label{fig:wIF_conv_M_photon_group}
\end{figure}
The time evolution of the ionisation-front width $w_\mathrm{i-front}$ is then shown in Fig.~\ref{fig:wIF_conv_M_photon_group} for $M=6,12$ and $24$. 
Up to $t\sim15\,\rm Myr$, the ionisation-front widths are nearly identical for all values of $M$. At later times, minor discrepancies of a few per cent begin to emerge between the simulations. Although small differences may persist in the formal limit $M\rightarrow\infty$, the overall front evolution remains well converged over the timescales most relevant to this work.

\subsubsection{PDF convergence test}
\begin{figure}
    \centering
    \includegraphics[width=1.\linewidth]{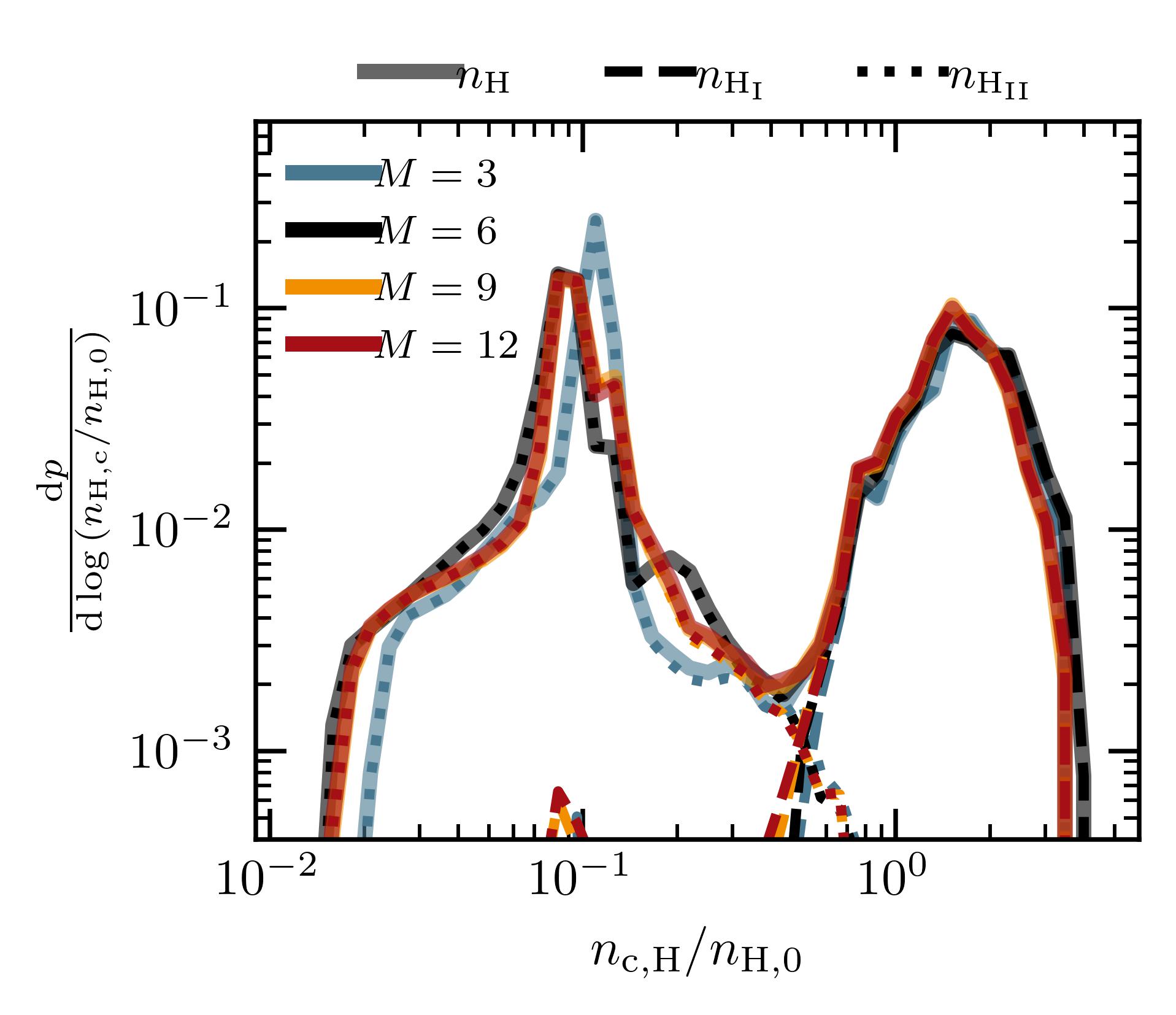}
    \caption{Cold-gas hydrogen number density PDFs weighted by number density and normalised to $n_{\rm H,0}$ at $t=5\,\rm Myr$ for a test simulation in the rocket-effect regime. The plain, dashed, and dotted lines show total, neutral, and ionised hydrogen, respectively. The light blue, black, orange and red lines correspond to $M=3,6,9$, and $12$, respectively.}
    \label{fig:PDF_conv_M_photon_group}
\end{figure}

Prior to the main simulation suite, we performed convergence tests for a representative simulation in the rocket-effect regime with $n_\mathrm{H,0}\sim 30\,\rm cm^{-3}$ and $F_\mathrm{q}\sim 10^{9.2} \, \rm cm^{-2}\,s^{-1}$. The simulations were evolved for $t=5\, \rm Myr$, and the resulting cold gas PDFs of $n_\mathrm{H}$,$n_\mathrm{H_I}$, and $n_\mathrm{H_{II}}$ are shown in Fig.~\ref{fig:PDF_conv_M_photon_group} for $M=3,6,9$ and $12$.
For $M=3$, the PDF peak associated with ionised gas is shifted towards slightly higher densities compared to the higher-$M$ simulations. In this case, the coarse discretisation of the SED smooths the distribution of photon energies and slightly modifies the thermal structure of the ionised gas. For $M\geq6$, the differences become relatively minor, indicating that the additional penetration and heating produced by harder photons do not substantially alter the thermodynamical evolution of the cloud.
\\
\\
Overall, the 1D tests suggest that $M=12$ may provide a more fully converged representation of the ionisation-front structure. However, in both the 1D and 3D simulations, the differences between $M=6$ and higher spectral resolutions remain modest and do not alter our results on the evolution of the clouds. We therefore consider $M=6$ to provide a reasonable compromise between physical accuracy and computational cost. In addition, the 1D simulations show that the ionisation front remains confined within the cloud throughout the evolution.

\section{Hydrogen lines effective rates}\label{app:eff_rates}
We describe the computation of the effective rates for Ly$\alpha$ and H$\alpha$.
\subsection{Effective recombination rates}
Effective recombination rates for hydrogenic line transitions are provided by \cite{Martin1988} and \cite{Pengelly1964} for the temperature ranges $78 < T < 2\times10^4 \, \mathrm{K}$ and $10^3 \lesssim T \lesssim 10^5 \, \mathrm{K}$, respectively.
For Ly$\alpha$, solely the 1s-2p transition contributes directly to the emission of a Ly$\alpha$ photon, meaning that the effective Ly$\alpha$ recombination coefficient is equal to the effective recombination coefficient to the 2p state. This is tabulated in the given references and already incorporates electron cascades from higher levels. For H$\alpha$, recombinations leading to all states 3s, 3p, and 3d are included. While in our main calculations (Sect.~\ref{sec:sub:emission}), we only consider case B, we also performed coefficient fits for case A. In this case, a branching ratio of 0.1183 is applied to the 3p state, to exclude the 3p-1s transitions that result in a Ly$\beta$, rather than H$\alpha$ photons\citep{Langen2023}.
Similarly to \cite{Cantalupo2008,Dijkstra2014}, we compute fitting formulae for the effective recombination rates of Ly$\alpha$ and H$\alpha$ of the form,
\begin{equation} \label{eq:H_rates2}
\alpha_{\mathrm{eff}} = 10^{-14}\times 10^{ C_3\tau^3 + C_2\tau^2 + C_1\tau + C_0} \, \rm cm^3 \, s^{-1},
\end{equation}
with the temperature parameter $\tau\equiv \log_2\left[ T / \left( 10^4 \, \rm K\right) \right]$, and $C_{i}$ the polynomial coefficients.
\begin{figure}
    \centering
    \includegraphics[width=1.\linewidth]{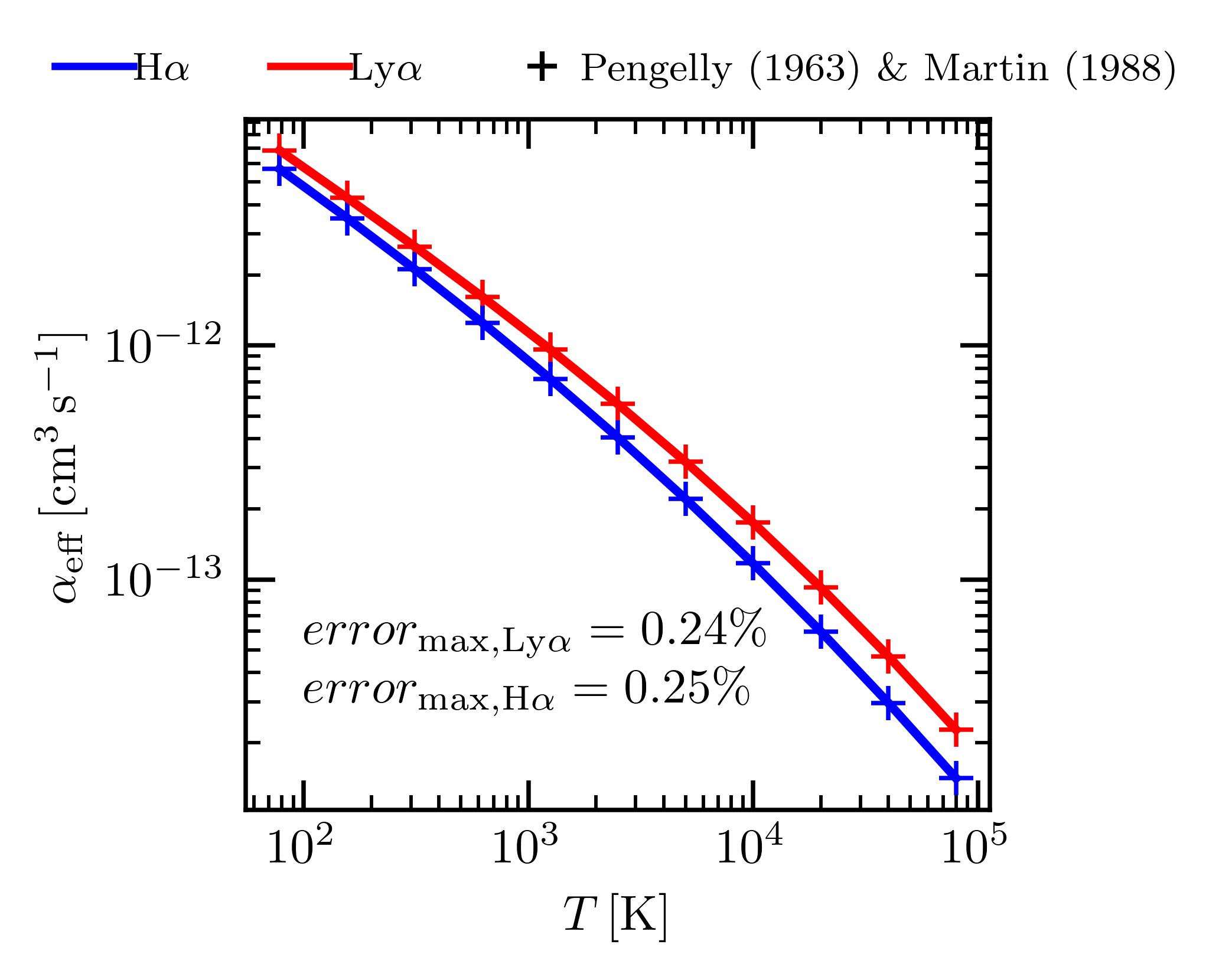}
    \caption{Effective recombination rate fits (lines) and tabulated values (crosses) for both Ly$\alpha$ and H$\alpha$ lines (Case B). }
    \label{fig:Ha_Lya_fits}
\end{figure}
The polynomial coefficients for case B are listed in  Table~\ref{tab:eff_rec_cst} for both Ly$\alpha$ and H$\alpha$ lines, and the fits are shown in Fig.~\ref{fig:Ha_Lya_fits}. The fits are valid for temperatures $78\,\mathrm{K} < T \lesssim 10^5 \,\mathrm{K}$, with maximum relative errors of approximately $0.25\,\%$ for both lines and mean relative errors of $0.09\,\%$ and $0.12\,\%$ for H$\alpha$ and Ly$\alpha$, respectively.

\begin{table}
\centering
\caption{Polynomial coefficients for the effective recombination rates (Case~B and Case~A) fitted from the tabulated values in \cite{Pengelly1964,Martin1988}.
}
\label{tab:eff_rec_cst}
\begin{tabular}{l|cc}
\hline
Coefficients & H$\alpha$ & Ly$\alpha$ \\
\hline
Case B &$\,$  &$\,$  \\
$C_3$&$-1.55783154\times 10^{-4}$&$-3.04114904\times 10^{-4}$\\
$C_2$&$-7.12756052\times 10^{-3}$&$-7.97320790\times 10^{-3}$\\
$C_1$&$-2.83292621\times 10^{-1}$&$-2.68368726\times 10^{-1}$\\
$C_0$&$1.06803625$&$1.24235523$\\
\hline
Case A &$\,$  &$\,$ \\
$C_3$&$-3.68065035\times 10^{-5}$&$-3.03381054\times 10^{-4}$\\
$C_2$&$-7.54973413\times 10^{-3}$&$-8.13840064\times 10^{-3}$\\
$C_1$&$-3.08762982\times 10^{-1}$&$-2.71325918\times 10^{-1}$\\
$C_0$&$0.890032905$&$1.22052631$\\
\hline
\end{tabular}
\end{table}
The coefficients for case~B and case~A (even though case~A  ones are not used) are shown in Table~\ref{tab:eff_rec_cst}. They are valid on the same temperature range and have mean relative errors of $0.56\,\%$ and $0.12\,\%$ for H$\alpha$ and Ly$\alpha$, respectively.

\subsection{Effective collisional rates}
The effective collisional rates for hydrogen transitions are provided through numerical fits from \cite{Giovanardi1987} and the corrected rates of \cite{Giovanardi1989}. They are valid for temperatures $2\times 10^3\,\mathrm{K}<T<5\times 10^5\, \mathrm{K}$.
For case B, transitions leading to the emission of a Ly$\alpha$ photon are primarily the 1s-2p transition, as well as higher-level transitions that subsequently cascade to 2p, such as 3s, 3d, 4p, and 4f. Owing to the small contribution of higher-level transitions, we restrict our calculation to levels up to 4f for Ly$\alpha$.
For H$\alpha$, in addition to the 3s, 3p and 3d transitions, we also consider 4s, 4p, 4d and 4f.
\begin{figure}
    \centering
    \includegraphics[width=1.\linewidth]{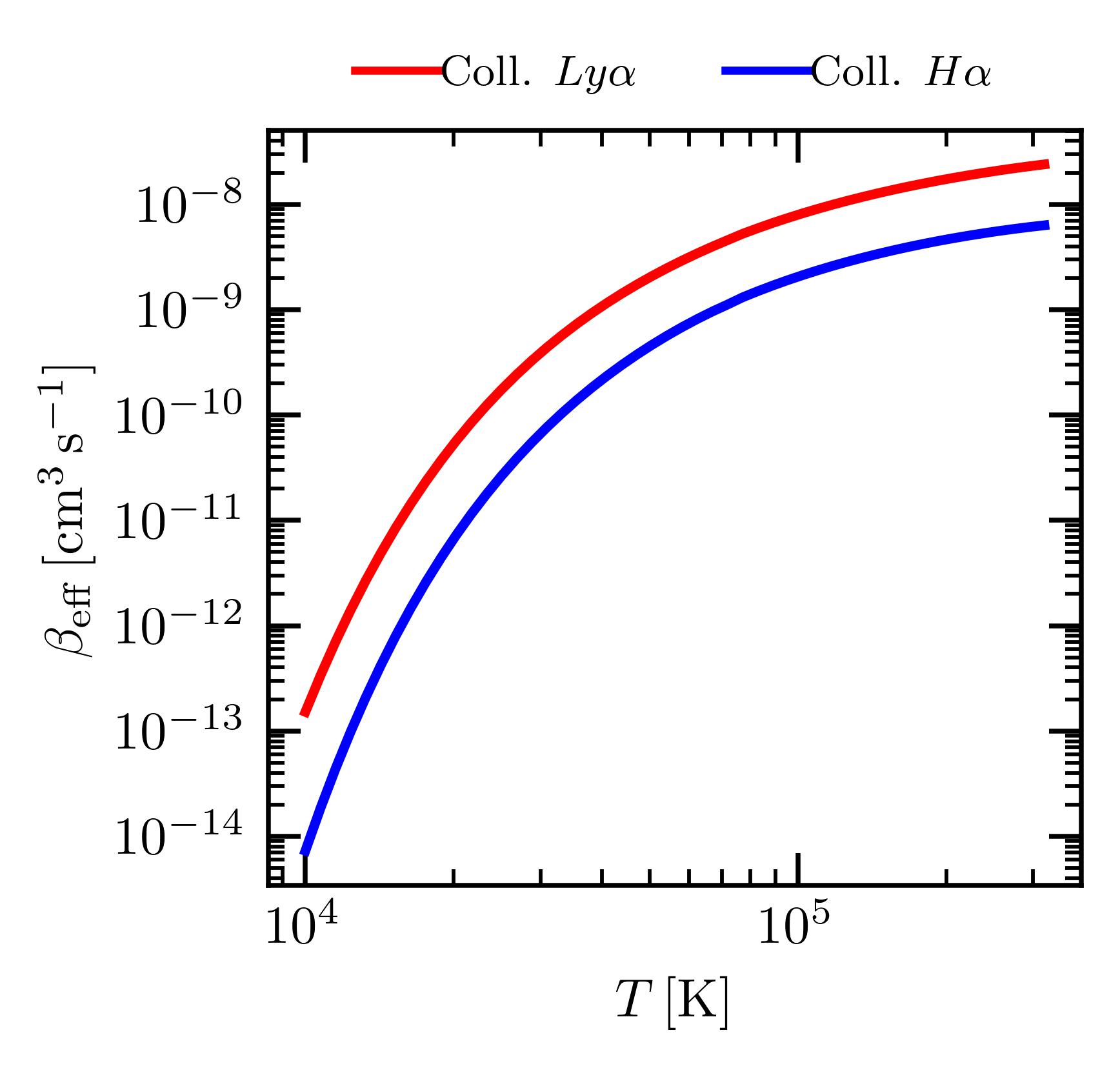}
    \caption{Effective collisional rates fits for both Ly$\alpha$ and H$\alpha$ lines (Case B from \cite{Giovanardi1987,Giovanardi1989}). }
    \label{fig:Ha_Lya_col_fits}
\end{figure}
The resulting collisional effective rates are shown in Fig.~\ref{fig:Ha_Lya_col_fits}. The rates decrease sharply for temperatures below $\sim 10^4 \,\mathrm{K}$. The ratio $\beta_{\mathrm{eff,Ly\alpha}} /\beta_{\mathrm{eff,H\alpha}} $ also increases significantly below $\sim 10^4 \,\mathrm{K}$, reaching values of $500$, $20$, and $3.8$ at $T = 4\times10^3\,\mathrm{K}$, $10^4\,\mathrm{K}$, and $10^{5.5}\,\mathrm{K}$, respectively. This ratio increases by a factor of 5.4 when expressed in energy units, owing to the different photon energies of the two lines.

\section{Alternative fits}\label{app:fits4obs}
\begin{figure}
    \centering
    \includegraphics[width=1.\linewidth]{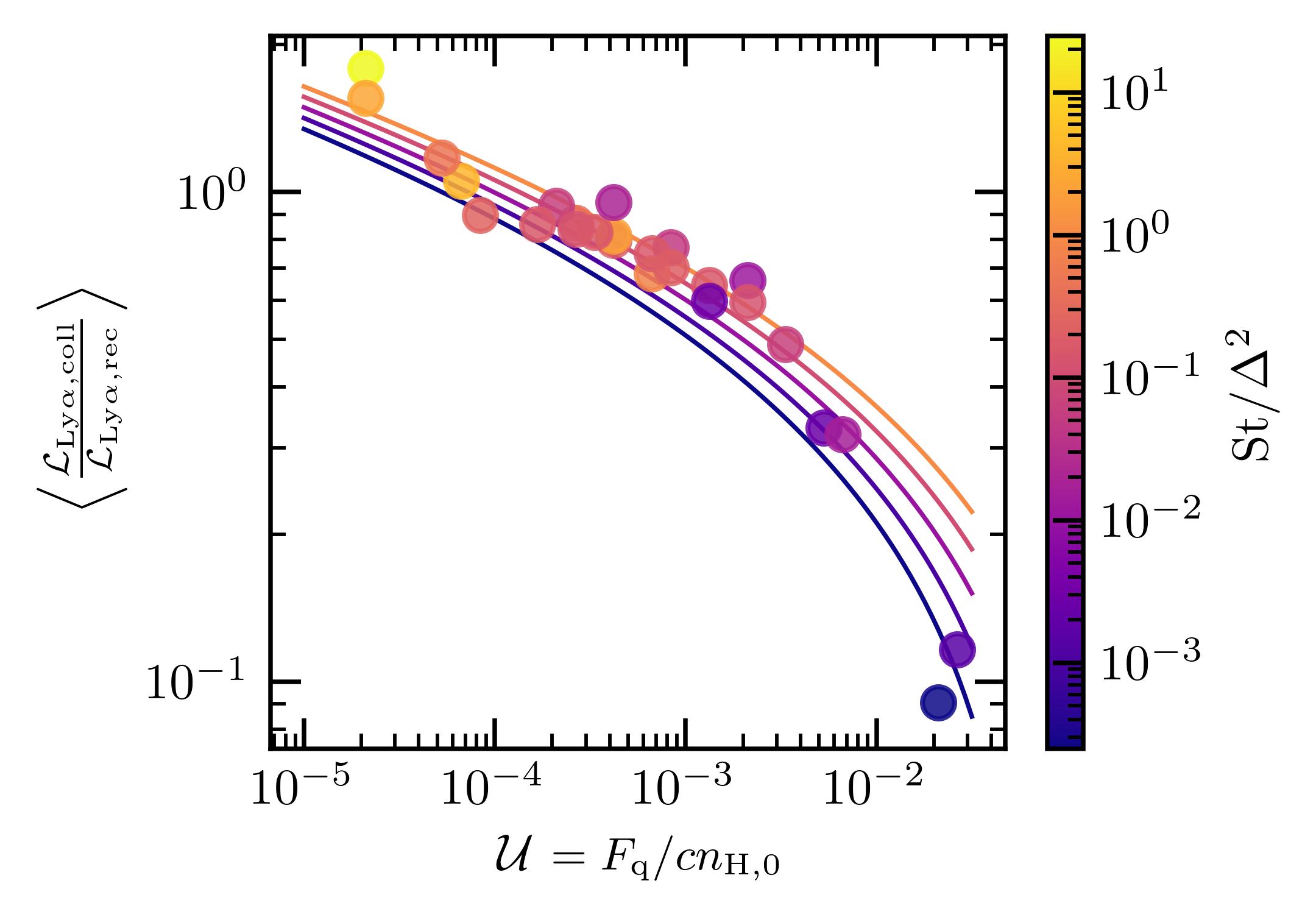}
    \\
    \includegraphics[width=1.\linewidth]{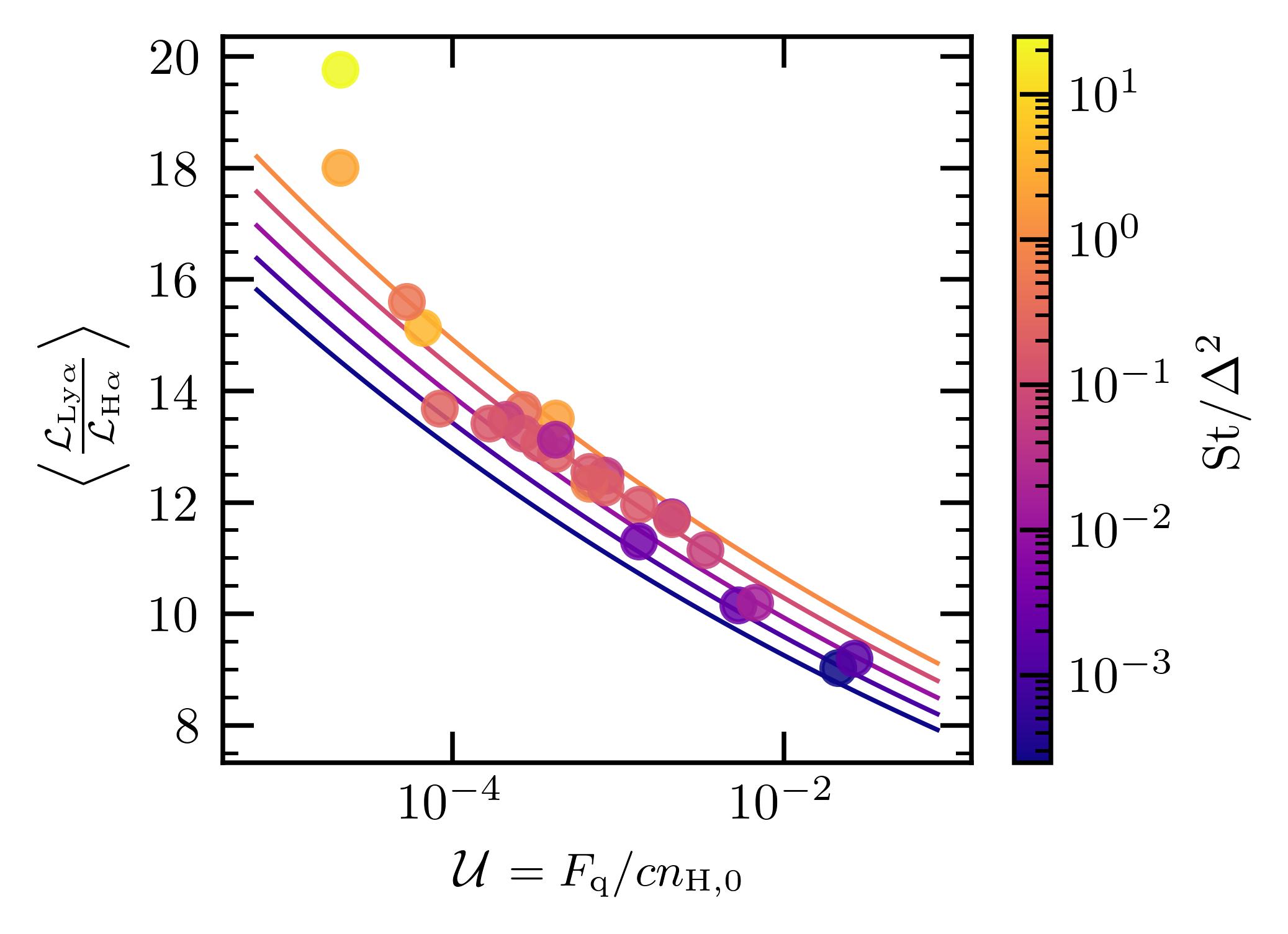}
    \caption{Ly$\alpha$ collisional to recombination luminosity ratio (top panel) and Ha$\alpha$ to Ly$\alpha$ luminosity ratio (bottom panel) as a function of the ionisation parameter $\mathcal{U}$ (x-axis) and the Strömgren number of the PIR $\mathrm{St}/\Delta^2$. The lines show the resulting fits using Eqs.~\ref{eq:r_Lya_rec_fit2}-~\ref{eq:r_Lya_Ha_fit2}.}
    \label{fig:U_fits}
\end{figure}

Following the discussion in Section~\ref{sec:sub:emission}, we here report our fits for the ratio of recombination to total Ly$\alpha$ luminosity, and for the Ly$\alpha$ to H$\alpha$ line ratio. The fits are expressed in terms of the ionisation parameter $\mathcal{U}=F_{\mathrm{q}}/n_{\mathrm{H,0}}c$ and Strömgren number for the PIR $\mathrm{St}/\Delta^2$.
This gives,
\begin{equation} \label{eq:r_Lya_rec_fit2}
\left\langle\frac{\mathcal{L}_{\mathrm{Ly\alpha,rec}}}{\mathcal{L}_{\mathrm{Ly\alpha}}}\right\rangle  \sim 1.14 \left(\frac{\mathrm{St}}{\Delta^2}\right)^{ -0.013 } \mathcal{U}^{0.0957},
\end{equation}
and
\begin{equation} \label{eq:r_Lya_Ha_fit2}
\left\langle\frac{\mathcal{L}_{\mathrm{Ly\alpha}}}{\mathcal{L}_{\mathrm{H\alpha}}}\right\rangle  \sim 7.602 \left(\frac{\mathrm{St}}{\Delta^2}\right)^{ 0.0153 } \mathcal{U}^{-0.0733},
\end{equation}
which yields a mean relative error of $0.5\,\%$ and $0.28\,\%$, respectively. To mention, the expression above should be capped by their maximum values, $1$ and $6.6-8.3$ for our temperature range, respectively.
We show the advantage of using the ionisation parameter $\mathcal{U}$ in Fig.~\ref{fig:U_fits} by plotting the Ly$\alpha$ collisional to recombination luminosity ratio, and the Ha$\alpha$ to Ly$\alpha$ luminosity ratio.
Not expressing the ratio in terms of total Ly$\alpha$ luminosity as in Fig.~\ref{fig:H_lum} and using the ionisation parameters $\mathcal{U}$ seems to drastically reduce the scatter and the dependency on $\mathrm{St}$. However, a counterpart of the fits is that $\mathrm{St}/\Delta^2$ and $\mathcal{U}$ are not orthogonal and thus fits at low $\mathcal{U}$ (e.g. $\sim 10^{-4}$) are only valid for high values of $\mathrm{St}/\Delta^2$ (e.g. $\sim 1$).

\section{Derivation of the analytical model for the halo}\label{app:model_halo}
In this appendix, we detail the formalism to derive the mean and global properties of an ensemble of clouds intersected along a line of sight.
\subsection{Initial assumptions}
Following Sect.~\ref{sec:discussion}, the cloud population is characterised by a radial density profile and a fixed volume filling factor, 
\begin{equation}\label{eq:app:ini_summary}
    n_{\mathrm{H,c}}\left(r\right) \sim n_0\left(\frac{r}{r_0}\right)^{-a_{\mathrm{n}}}, \, f_{\mathrm{V}}= 0.01, 
\end{equation}
where $r_0=10\, \kpc$ is chosen as an initial radius defining the CGM, $n_0$ is the hydrogen number density at $r_0$, and $a_{\mathrm{n}}=1.25$ the slope. 
In addition, while the integral Strömgren number is independent of the cloud size, we also define a characteristic cloud radius proportional to the local Jeans length to derive the mean Strömgren number,
\begin{equation}\label{eq:app:ini_summary2}
    r_{\mathrm{c}}\left(r\right) = \frac{bA_{\mathrm{J}}}{2}n_{\mathrm{H,c}}\left(r\right)^{-1/2},
\end{equation}
where $b$ is the proportionality factor strictly smaller than unity, and $A_{\mathrm{J}}=c_{\mathrm{s,c}}\sqrt{\pi X/m_{\mathrm{p}}G }$ is a constant.

\subsection{Derivation of the cloud distribution along a ray}
We consider a thin spherical shell at radius $r$ with volume $\mathrm{d}V_{\mathrm{shell}} = 4\pi r^2 \mathrm{d}r$. The total volume of cold gas within this shell can be expressed either in terms of the volume filling factor or as the sum over individual clouds, leading to
\begin{equation}\label{eq:app:Vcol_shell}
V_{\mathrm{cloud}}\left(r\right)\mathrm{d}N_{\mathrm{cloud,shell}}\left(r\right)=  f_{\mathrm{V}} 4\pi r^2\mathrm{d}r.
\end{equation}
This yields the cloud volume-distribution,
\begin{equation}\label{eq:app:d_V}
d_{\mathrm{V}}\left(r\right) \equiv \frac{\mathrm{d} N_{\mathrm{cloud,shell}}\left(r\right)}{\mathrm{d} V_{\mathrm{shell}}\left(r\right)}=\frac{3f_{\mathrm{V}}}{4\pi r_{\mathrm{c}}\left(r\right)^3},
\end{equation}
which here corresponds to the cloud volumetric number density at a given halo-centric radius $r$.
To relate this 3D distribution to the number of clouds intersected along a line of sight, we consider the expected number of line-cloud intersections. Each cloud presents a geometrical cross-section $\sigma(r) = \pi r_{\mathrm{c}}(r)^2$, such that the line distribution is obtained as
\begin{equation}\label{eq:app:N_int}
d_{\mathrm{l}}\left(r\right) \equiv \frac{\mathrm{d} N_{\mathrm{cloud,line}}}{\mathrm{d} r} = \sigma\left(r\right) d_{\mathrm{V}}\left(r\right).
\end{equation}
Substituting Eq.~\ref{eq:app:d_V}, one obtains
\begin{equation}\label{eq:app:d_l}
d_{\mathrm{l}}\left(r\right)= \frac{3f_{\mathrm{V}}}{4 r_{\mathrm{c}}\left(r\right)}.
\end{equation}

\subsection{Integral quantities}
Having defined the line distribution $d_{\mathrm{l}}$, we may now derive integrated quantities along a line of sight. The zeroth moment yields the cumulative number of clouds intersected by the line of sight up to radius $r$,
\begin{equation}\label{eq:app:Nc_l}
    N_{\mathrm{c,l}}\left(r\right) =\int^r_{r_0} d_{\mathrm{l}}\left(r'\right) \,\mathrm{d}r'= \frac{3r_0n_0^{1/2}f_{\mathrm{V}}}{\left(2-a_{\mathrm{n}}\right)bA_{\mathrm{J}}}\left(\left(\frac{r}{r_0}\right)^{1-a_{\mathrm{n}}/2}-1\right).
\end{equation}
The associated line-ensemble integrated and mean line quantities, $\mathrm{St}_{\mathrm{l}}$ and $\left\langle \mathrm{St} \right\rangle_{\mathrm{l}}$, are given by
\begin{equation}\label{eq:app:St_l}
     \mathrm{St}_{\mathrm{l}}\left(r\right) = \int^r_{r_0} \mathrm{St}\left(r'\right) d_{\mathrm{l}}\left(r'\right) \,\mathrm{d}r' = \frac{6\pi\alpha^{\mathrm{B}}_{\mathrm{H}}n_0^2r_0^3f_{\mathrm{V}}}{\left(3-2a_{\mathrm{n}}\right)\dot{N}_{\mathrm{ph}}}\left(\left(\frac{r}{r_0}\right)^{3-2a_{\mathrm{n}}}-1\right),
\end{equation}
and
\begin{equation}\label{eq:app:St_ml}
     \left\langle \mathrm{St} \right\rangle_{\mathrm{l}}\left(r\right) = \frac{1}{N_{\mathrm{c,l}}\left(r\right)} \int^r_{r_0} \mathrm{St}\left(r'\right) d_{\mathrm{l}}\left(r'\right) \,\mathrm{d}r'=\frac{\mathrm{St}_{\mathrm{l}}\left(r\right)}{N_{\mathrm{c,l}}\left(r\right)}.
\end{equation}
The quantity $\mathrm{St}_{\mathrm{l}}$ characterises the cumulative interaction of radiation with the ensemble of clouds along the line of sight, under the simplifying assumption that the incident flux is not significantly attenuated within individual clouds. In this sense, it provides an upper bound on the transmitted flux and a corresponding lower bound on $\mathrm{St}_{\mathrm{l}}$.
The parameter $\langle \mathrm{St} \rangle_{\mathrm{l}}$ represents the mean value of $\mathrm{St}$ across all intersected clouds.

Physically, $\mathrm{St}_{\mathrm{l}}$ plays a role analogous to the single-cloud quantity $\mathrm{St}$. If $\mathrm{St}_{\mathrm{l}}<1$, the radiation field is capable of fully ionising and penetrating the cold gas throughout the halo. Conversely, if $\mathrm{St}_{\mathrm{l}}>\Delta^2$, the cold CGM becomes effectively radiation-shielded, and the radiation cannot propagate beyond the radius $r$. In the rocket-effect regime, $1<\mathrm{St}_{\mathrm{l}}<\Delta^2$, an ionisation front is expected to form and advance through multiple clouds at its characteristic front velocity, rather than at the speed of light.

\section{Quasar and cold inflows}\label{app:St_s}
Massive galaxies at redshift $z \gtrsim 1$ are thought to be sustained by cold gaseous streams accreting from the IGM \citep{Fardal2001,Dekel2006,Waterval2025}. Though covering only a small volume fraction of the halo \citep{Faucher-Giguere2011}, these streams still constitute its densest component.
We thus investigate the implication of our theoretical framework on an idealised inflow model \citep[see also][for the mathematical description]{Aung2024,Ledos2024b}. The investigation is straightforward, as the cold infow model from \citet{Dekel2013}, then refined by \citet{Mandelker2020b}, gives lower and upper limit values for the cold stream hydrogen number-density profile $n_{\mathrm{H},\stream}(r)$ and the radial stream profile $r_{\stream}(r)$. The stream radius profile we assume follows \citet{Aung2024}.

\subsection{Impact of a ray propagating radially through a cold stream}
Focusing on one ray propagating solely inside the stream and not a statistical ensemble, the formalism described in Appendix~\ref{app:model_halo} then reduces to,
\begin{equation}\label{eq:app:St_ls}
     \mathrm{St}_{\mathrm{l,stream}}\left(r\right) =\int^r_{r_0} \frac{\alpha^{\mathrm{B}}_{\mathrm{H}}n_{\mathrm{H,\stream}}\left(r'\right)^2}{F_{\mathrm{q}}\left(r'\right)} \,\mathrm{d}r'.
\end{equation} 
Using the fiducial bound from \citet{Mandelker2020b} for the hydrogen number density normalisation $n_0$, we plot the results in Fig.~\ref{fig:St_l}.

\subsection{Impact of one ray crossing a cold stream from the side}
Assuming that the cold stream has an impact parameter allowing a ray to roughly cross it perpendicularly, we then obtain the Strömgren number for the cold stream at a given $r$ as, 
\begin{equation}\label{eq:St__s}
    \mathrm{St}\left(r\right) = \frac{2r_{\stream}\left(r\right)\alpha^{\mathrm{B}}_{\mathrm{H}}n_{\mathrm{H,\stream}}\left(r\right)^2}{F_{\mathrm{q}}\left(r\right)},
\end{equation}
where we assume $\alpha^{\mathrm{B}}_{\mathrm{H}}$ to be constant as it is defined at the temperature of the ionised cold gas $T_{\mathrm{i}}$. Here, $\mathrm{St}(r)$ characterises radiation that intersects the cold stream locally at a halo-centric radius $r$.
\begin{figure}
    \centering
    \includegraphics[width=1.\linewidth]{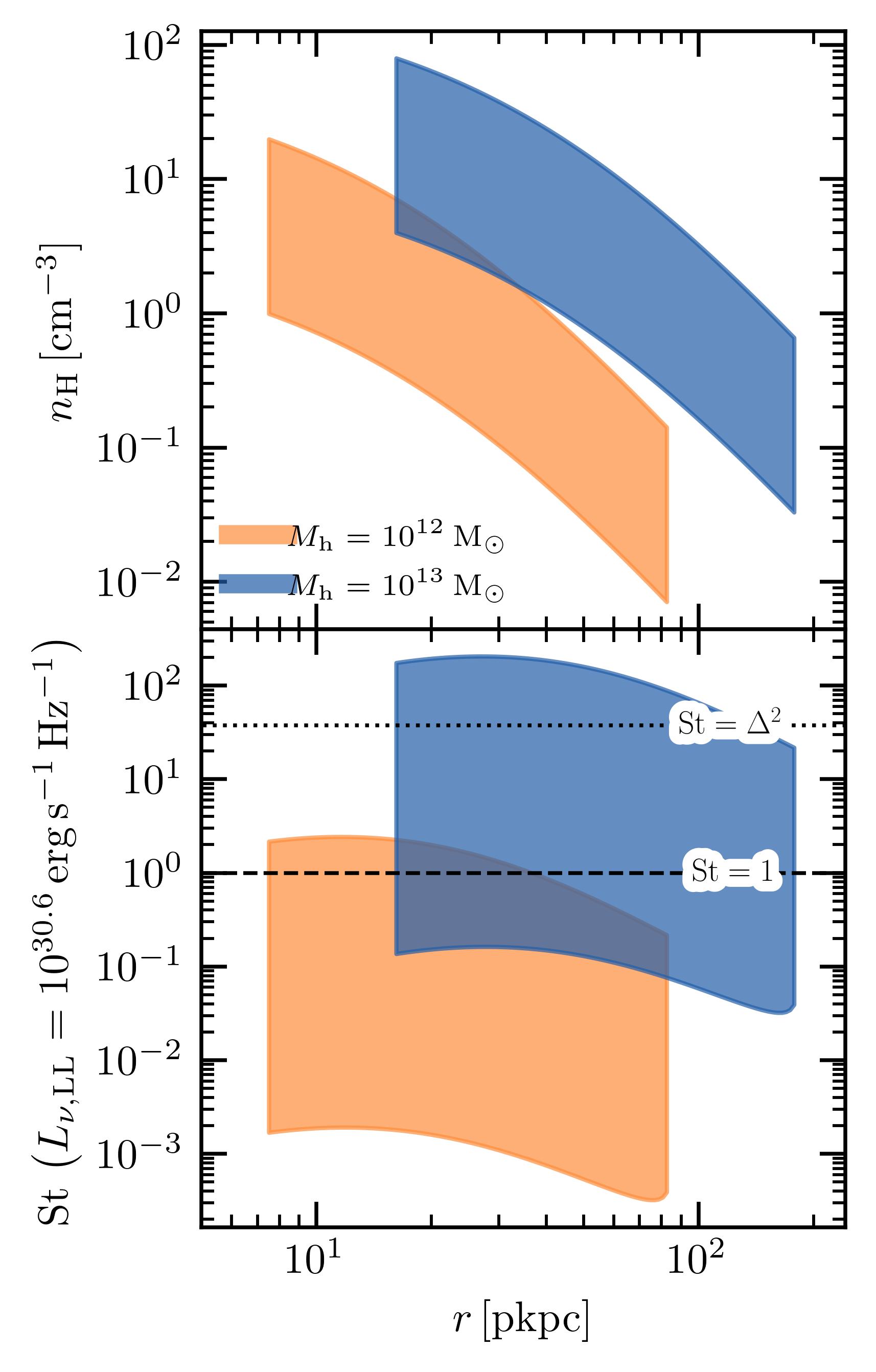}
    \caption{Radiation impact on cold inflows. The cold stream toy model assumes a redshift $z=3$ and two halo masses of $M_{\mathrm{h}}=10^{12}$and $10^{13}\,\rm \Msun$. The quasar specific luminosity is fixed with $L_{\mathrm{\nu,LL}}=10^{30.6}\,\rm  erg\, s^{-1}\, Hz^{-1}$. The radial profiles are plotted from $10\%$ to $110\%$ of the virial radius of their associated halo. {\it Top panel:} Cold stream hydrogen number density. {\it Bottom panel:} $\mathrm{St}$ parameter of the stream, assuming that the radiation crosses the cold flow solely at a halo radius $r$.}
    \label{fig:St__prof_s}
\end{figure}
\\
Figure~\ref{fig:St__prof_s} shows the resulting radial profiles of $\mathrm{St}$ for halo masses of $M_{\mathrm{h}}=10^{12}$ and $10^{13}\,\rm \Msun$ at $z=3$, assuming quasar specific luminosity $L_{\mathrm{\nu,LL}}=10^{30.6}\,\rm  erg\, s^{-1}\, Hz^{-1}$ At this redshift, the mean gas density in the Universe is relatively high, leading to dense cold streams already upon entry into the halo. The order-of-magnitude difference in halo mass translates into an approximately tenfold increase in stream density at fixed halo-centric radius, and hence into a factor of $\sim100$ variation in $\mathrm{St}(r)$, given its quadratic dependence on density. Accordingly, a cold stream within a $10^{12}\,\Msun$ halo is expected to lie in the optically thin regime, whereas streams in $10^{13},\Msun$ haloes are more likely to reside in the rocket-effect.
Given the limitation described in Sect.~\ref{sec:discussion}, an extended quantitative assessment of cold inflow survival would require additional treatment. Nevertheless, in qualitative terms, cold streams appear predisposed to enter the rocket-effect regime, wherein they potentially efficiently emit their Ly$\alpha$ photon via both recombination and collisional processes.

\section{Estimate of the cometary globule acceleration}\label{app:acc_CG}
We derive a simple estimate of the acceleration of a cometary globule in the rocket-effect regime, driven by the recoil from photo-evaporating gas in the photoionised region (PIR). This regime is reached once the neutral gas ahead of the ionisation front can no longer be significantly compressed, and the remaining cold clump is accelerated by the momentum flux of the evaporating ionised gas.

Following \citet{Nakatani2019}, the initial acceleration of the cometary globule may be approximated as
\begin{equation}\label{eq:app:a0}
    a_{\mathrm{CG}} = \frac{1}{M_{\mathrm{GC}}}\frac{\sigma_{\mathrm{GC}}}{2}\rho_bv_b^2,
\end{equation}
where $M_{\mathrm{GC}}$ and $\sigma_{\mathrm{GC}}$ are the mass and effective cross-section of the globule, and $\rho_b$ and $v_b$ are the density and velocity of the ionised gas at the base of the ionisation front. The evaporating gas is assumed to flow at approximately the ionised sound speed $c_\mathrm{i}$ \citep{lefloch_cometary_1994}, which we compute from the adopted ionised gas temperature $T_\mathrm{i}$. This choice of velocity also directly leads to the dynamical pressure at the base of the front being equal to the thermal one, i.e. $\rho_bc_\mathrm{i}^2 = \rho_bk_\mathrm{b}T_\mathrm{i}/m_\mathrm{i}$, with $m_\mathrm{i}\sim 0.67m_\mathrm{p}$ the atomic mass in of the fully ionised gas. From the middle panel of Fig.\ref{fig:nH_pdf}, we estimate $\rho_b \sim 0.4\rho_{\mathrm{c,0}}$.

We further express the globule properties in terms of the initial cloud quantities,
\begin{equation}\label{eq:app:rhoGC}
    \rho_{\mathrm{GC}} = \rho'\rho_{\mathrm{c,0}} = \rho'\frac{m_p n_{\mathrm{H,0}}}{X},
\end{equation}
and
\begin{equation}\label{eq:app:MGC}
    M_{\mathrm{GC}} = m'M_{\mathrm{c,0}}=m'\frac{m_p n_{\mathrm{H,0}}}{X}\frac{4\pi r_{\mathrm{c,0}}^3}{3},
\end{equation}
where $\rho'$ and $m'$ denote the globule density and mass in units of the initial cloud values.
Combining these relations yields
\begin{equation}\label{eq:app:a1}
    a_{\mathrm{CG}} = \frac{3}{8r_{\mathrm{c,0}}}\frac{1}{m'}\left(\frac{m'}{\rho'}\right)^{2/3}\, 0.4c_\mathrm{i}^2.
\end{equation}

\begin{figure}
    \centering
    \includegraphics[width=1.\linewidth]{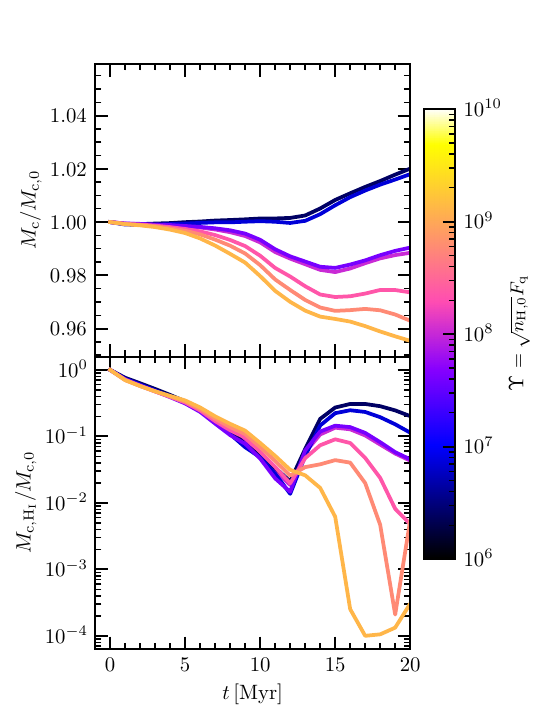}
    \caption{Cold gas total mass (top) and \ion{H}{I} cold gas mass (bottom), both normalised by the initial cloud total mass. The simulations are fixed at $\mathrm{St} \sim 8$ (rocket-effect regime) with a varying ionisation strength parameter $\Upsilon$.}
    \label{app:St_prof_M}
\end{figure}
Figure~\ref{app:St_prof_M} shows the evolution of the total cold gas mass $M_{\mathrm{c}}$ and the neutral component $M_{\mathrm{c,H_I}}$ for simulations simulations are at a fixed $\mathrm{St} \sim 8$ (rocket-effect regime) with varying ionisation strength parameter $\Upsilon$. The total cold mass remains approximately constant throughout the evolution. By contrast, $M_{\mathrm{c,H_I}}$ decreases monotonically until $t \sim 12\,\mathrm{Myr}$, tracing the propagation of the ionisation front, the formation of the cometary globule around $t\sim 6\,\rm Myr$ (see Fig.\ref{fig:evo_cloud_ups}) and its initial acceleration. 
At $t\gtrsim 12\,\rm Myr$, the globule follows a more complex evolution where it either fragments or ionises. Its evolution depends on $\Upsilon$: for large values, the globule is gradually ionised, while for $\Upsilon \lesssim 10^{8.5}$ it first fragments into multiple clumps along its shadow, which are then partly ionised (see Fig.\ref{fig:evo_cloud_ups}), leading to overall mass decreasing back to approximately its initial one.
Although the long-term evolution is not fully captured due to the limited box size and simulation time, we correctly capture the initial formation of the cometary globule for which Fig.~\ref{app:St_prof_M} suggests a characteristic mass fraction $m' \sim 0.15$ at the initial cometary globule stage at $t\sim 6$. From Fig.~\ref{fig:nH_pdf}, we similarly estimate $\rho' \sim 2$.

Substituting these values into Eq.~\ref{eq:app:a1}, we obtain
\begin{equation}\label{eq:app:a2}
    a_{\mathrm{CG}} = 2.04 \left(\frac{50 \, \mathrm{pc}}{r_{\mathrm{c,0}}}\right)\left(\frac{T_{\mathrm{i}}}{10^{4.3} \,\mathrm{K}}\right) \quad \rm pc \, Myr^{-2},
\end{equation}
which provides a representative estimate for clouds in the rocket-effect regime with $\mathrm{St}\sim 8$.
Assuming constant acceleration and neglecting mass loss, the globule may be treated as a solid body over a timescale $t \sim 4\,\mathrm{Myr}$. This yields a characteristic displacement of $\sim 30\,\mathrm{pc}$ and a velocity of $\sim 8\,\mathrm{km\,s^{-1}}$, in good agreement with the displacement shown in our simulations.

For comparison, we estimate the gravitational acceleration from a Navarro–Frenk–White (NFW) halo,
\begin{equation}\label{eq:app:aNFW}
    a_{\mathrm{NFW}}(x) = \frac{GM_{\mathrm{h}}}{m(c)}\times\left( \frac{cx}{cx+1}-\log{cx+1}\right)x^{-2}R_{\mathrm{v}}^{-2},
\end{equation}
where $x\equiv rc/R_{\mathrm{v}}$ with $r$ is the halo radial coordinate, $G$ the gravitational constant and $m\left(c\right)=\log{\left(c+1\right)} - c/\left(c+1\right)$.
For a halo of mass $M_{\mathrm{h}}=10^{12}\,\mathrm{M_\odot}$, virial radius $R_{\mathrm{v}}\sim 75\,\mathrm{pkpc}$, and concentration $c=5$ at $z=3$, we find  $a_{\mathrm{NFW}} \sim 5.2$ and $0.8 \, \rm pc \, Myr^{-2}$ for $r=10$ and $75\,\rm pkpc$, respectively.
Equating $a_{\mathrm{NFW}} = a_{\mathrm{CG}}$ yields a characteristic radius $r \sim 33\,\mathrm{pkpc}$. Beyond this radius, the rocket-effect acceleration becomes comparable to or exceeds the gravitational pull, and may therefore potentially reduce, or even counteract, the free-fall motion of the cometary globule.

\end{appendix}

\end{document}